\documentclass{aa}  %

\usepackage{graphicx}
\usepackage{txfonts}
\usepackage{xcolor}

\begin{document}

   \title{The applicability of the JAGB method for measuring the distance 
          of galaxies subject to different metal enrichment rates}


\author{C. Gavetti\inst{1,2}, P. Ventura\inst{2}, F. Dell'Agli\inst{2}, L. Graziani\inst{3}, M. Correnti\inst{4,5}, C. Ventura\inst{2}, F. La Franca\inst{1}}

   \institute{Dipartimento di Matematica e Fisica, Università degli Studi Roma Tre, 
              via della Vasca Navale 84, 00100, Roma, Italy \and
              INAF, Observatory of Rome, Via Frascati 33, 00077 Monte Porzio Catone (RM), Italy \and
              Dipartimento di Fisica, Sapienza, Università di Roma, Piazzale Aldo Moro 5, 00185, Roma, Italy \and
              ASI-Space Science Data Center, Via del Politecnico, I-00133, Rome, Italy \and
              Istituto Nazionale di Astrofisica– Osservatorio Astronomico di Padova, Vicolo dell'Osservatorio 5, I-35122 Padova, Italy
              }

   \date{Received September 15, 1996; accepted March 16, 1997}


 \abstract
   {The JAGB method has been proposed over the last years as a possible distance
    indicator for the galaxies in the Local Group and beyond. The nature of
    the stars populating the J region, as well as the conditions of the star formation history and the structural properties of the galaxies
    for the straight application of this method, still need to be investigated.}
   {We assess the robustness of the JAGB method to measure the distance
   of galaxies, presenting a detailed theoretical analysis of the stars 
   populating the J region of the $\rm (J-K_S, J)$ diagram. The primary 
   goal of the present analysis is to identify the properties of the 
   corresponding J luminosity function (JLF) unaffected by the previous 
   history of the galaxies, in terms of the metal enrichment of the 
   interstellar medium.}
   {We use a population synthesis approach, based on extant stellar models of
   AGB evolution coupled self-consistently with the description of the
   dust formation process in the wind. This is done to build synthetic distribution of
   stars on the $\rm (J-K_S, J)$ diagram, and calculate the corresponding
   JLF of the stars in the J region. The simulations are reiterated for different hypothesis regarding
   the time variation of the metallicity of the interstellar medium, to
   study the sensitivity of the JLF shape to the velocity of
   the metal enrichment process of the galaxies.}
   {The statistics of the JAGB region is mostly determined by the stars
   formed between $\sim 6$ Gyr and $\sim 1$ Gyr ago, considering that all the stars
   formed during epochs external to this time window barely enter the
   JAGB box. The shape of the JLF changes significantly with the
   metal enrichment, and in particular the J magnitude peak, which
   change over a range of values extending over 0.3 mag. Conversely,
   the mean J magnitude $\rm M_J^{av}$ proves much less dependent of the previous
   history of the galaxy and thus a much more robust distance indicator:
   we find $\rm M_J^{av}=-6.2 \pm 0.05$ mag for all the cases investigated.
   The uncertainties connected
   to the still largely unknown process of mass loss suffered by low-mass
   stars during the evolution along the red giant branch are also commented.}
   {}


   \titlerunning{The JAGB method across different metallicities}
   \authorrunning{Gavetti et al.}
   \maketitle
%

\section{Introduction}
\label{intro}
The last decades have witnessed a growing interest in  
the stars evolving through the asymptotic giant branch
(AGB), owing to their efficiency as pollutants of the interstellar 
medium and the important role played in the chemical enrichment
of galaxies through their cosmic evolution history \citep{ciaki20,romano22}, including the formation of 
multiple populations in globular clusters \citep{ventura01}. 
Further interest in this class of objects arose when they
were recognized as among the most efficient dust manufacturers,
providing a significant contribution to the
dust production, also in the high-redshift Universe
\citep{valiante09, valiante11, mancini15, ginolfi18, aoyama18,graziani20,raffa24}.

Stellar modelers have recently included the description
of the dust formation process in the wind of cool stars, thus
making it possible to determine the dust production rate
(DPR) of stars, as they evolve along the AGB \citep{ventura12, 
ventura14, nanni13, nanni14}. These efforts have been dictated not
only by the need to infer the mineralogy and the amount of dust 
produced by AGB stars,
but also to understand the evolution of their spectral energy
distribution (SED), given the reprocessing of the
radiation released by the photosphere by the dust particles present
in the circumstellar envelope. The coupling of dust formation with
stellar evolution modeling is essential for the 
computation of the evolutionary tracks of stars of different mass and
metallicity across observational color-magnitude diagrams (CMDs) built with near-infrared (NIR) and mid-infrared (MIR)
filters. This approach was used to characterize the evolved stellar
population of the Magellanic Clouds (MCs) \citep{flavia14b, flavia15a, flavia15b, 
nanni16, nanni18, nanni19}, of a few Local Group (LG) galaxies
\citep{flavia16, flavia18, flavia19} and of Andromeda \citep{cla25}.

An accurate description of the evolution of AGB stars and of 
dust formation in their winds is essential more than ever, now that we have entered the James Webb Space Telescope (JWST) era.
Indeed, the JWST enables the detection of resolved stellar populations in 
galaxies well beyond the LG and provides an exceptional tool for studying 
AGB stars \citep{correnti25, bortolini25}. For a significant fraction of 
these systems, AGB stars represent important tracers for reconstructing 
their star formation histories (SFHs), due to their age sensitivity in 
the NIR \citep{Lee24b,bortolini24}.
In this regard, recent studies have outlined the potential offered by
the number counts of evolved stars in reconstructing the 
SFH of galaxies, a method so far applied to
the Milky Way satellites, based on HST data \citep{harmsen23}.

The proposal to use AGB stars, or at least a part of them, as distance 
indicators of galaxies, has recently been added to the themes discussed above
and has further underlined the need for an exhaustive understanding of 
the evolutionary properties of these objects. This is the so called 
JAGB method, first proposed by \citet{nikolaev00}, who suggested to use
the AGB stars located in a rectangular box of the $\rm (J-K_s, K_s)$ diagram of the LMC,
later called the "J region", as standard candles. An accurate calibration of the
JAGB method would lead to the measurement of the distances of galaxies that 
are found to be further away than those already determined using other methods.
Indeed, JAGB stars are significantly brighter on average than the tip of the red giant 
branch (TRGB), allowing distance determinations well beyond the TRGB limit. 
Compared to Cepheids, the JAGB method requires only a single epoch of observations 
and it can be applied to galaxies hosting stellar populations with ages between $\sim 200$ Myr and 1 Gyr, whereas Cepheids are confined to the disks of spiral and irregular systems. 

While the initial investigations were based on the MCs, following works
investigated the J luminosity function (JLF) of other galaxies, and confirmed
that the applicability of the JAGB method can be considered as general.
Important contributions in this field came from \citet{madore20} and 
\citet{freedman20}, who studied the JLF of 16 galaxies, finding that 
all the JLFs can be approximated with an approximately mono-modal,
slightly asymmetric gaussian function, peaking at $\rm M_J = -6.2$ mag,
with a small scatter of the order of 0.02 mag. A more recent and exhaustive
investigation was presented by \citet{freedman25}, who focused on the
JLF of 70 galaxies observed in the I band, and found little evidence
for invoking multi-modal fits. Further applications of this technique to 
other galaxies were presented by \citet{Lee24a} and \citet{Li25}.

\citet{magnus24} have recently refined the calibration of the 
JAGB method using Gaia data on the MCs, selecting a clean carbon star sample within $\rm 1.5 < J-K < 2.0$ mag and a 1.2 mag wide J window: they found a mean $\rm M_J$ of $\sim -6.25$ mag for the LMC and $\sim -6.18$ mag for the SMC. \citet{cla26} applied stellar evolution modeling of AGB stars, coupled 
with the description of dust formation,
to interpret the results obtained by \citet{magnus24}, and
successfully reproduced the JLF derived by \citet{magnus24} 
for the LMC and the SMC. The analysis by \citet{cla26} outlined that 
the population of the J region is composed of stars evolving through
a well-defined evolutionary phase during the AGB, which extends from the early phases following the C-star stage until the amount of carbon dust in the circumstellar envelope becomes large enough to shift the SED entirely to the MIR region, so that
they move to the $\rm J-K > 2.0$ mag region of the CMD. An important side
result obtained by \citet{cla26} is that the possibility for stars to
evolve into the J region, and then the time spent there, is
extremely sensitive to the progenitor's mass and to the metallicity.
Because stars of different mass and metallicity nowadays
evolving along the AGB formed in different epochs, the results
obtained by \citet{cla26} show that the numerical
consistency and the distribution of the stars in the J region
are tightly connected to the previous history of galaxies, in terms 
of the metal enrichment of the interstellar medium and of the SFH.

A more general approach is required to test the applicability of the JAGB method to different galaxies from the LMC and SMC,
where no a priori choice on the SFH is made: this is the
only way to explore the conditions under which the 
JAGB method can be universally applied to measure
the distances of galaxies. The present research project aims
at a critical evaluation of the applicability of the JAGB
method, in relation to the type of galaxies investigated.
One of the objectives of this study is to test whether some
properties of the JLF of the stars, e.g. the mean and the peak 
J magnitudes, are independent of the previous history of metal enrichment of the galaxies.

In this work, we use the same approach adopted in \citet{cla26} to 
produce synthetic distributions of stars, to study in detail 
the population of the J region. We focus on the role played by the 
metal enrichment, by varying the rate at which the metallicity of
the interstellar medium increases over time. While the evolution of different galaxy types is generally characterized by different SFH, for the sake of simplicity, in the present analysis we assume a constant star formation rate (SFR)
during the various epochs; a more general analysis, where changes 
in the SFR are taken into account for different class of galaxies, will be the subject of a forthcoming paper. Part of the study is devoted to discuss
the relevance of the assumptions regarding the mass loss experienced by low-mass
stars during the red giant branch (RGB) evolution: this point
is often neglected in the literature, despite it was shown to
be crucial for the interpretation of the evolved stars distribution in the galaxies on the observational diagrams \citep{cla25}, and to
heavily affect the number counts and the relative fraction of AGB over
RGB stars in the Milky Way satellites \citep{ventura26}.

This work is structured as follows: the physical and numerical
ingredients used for the simulations are described in Section
\ref{input}; in Section \ref{tracks} we discuss the general
properties of the AGB evolution of stars of different mass and
metallicity, and the time spent in the J region; the sensitivity
of the shape of the JLF to the timescale of the metal enrichment
of the galaxies is addressed in Section \ref{jlf}; Section \ref{disc}
is devoted to the analysis of the applicability of the J method
to infer the distance of galaxies; finally,
the conclusions are given in Section \ref{concl}.

\section{Physical and numerical input}
\label{input}
The present work is based on results obtained through a population synthesis approach. Artificial distribution of stars in the CMD of a given galaxy are simulated according to specific assumptions about their SFH that characterise the time from the galaxy’s formation to the present. As stated in the previous section, the main goal of the present investigation is the study of how the metal enrichment affects the population of
the J region of the CMD. Therefore, as a first approach, we assume a constant SFR,
while the metallicity of the gas increases with a timescale that is allowed
to vary in the different simulations. \\
It should be noted that proper assumptions on the SFH should be based first on the galaxy type under investigation \citep{ConroyReview13, MadauDickinsonReview14}, second on their physical properties, mainly their stellar mass \citep{Pacifici16} and finally their dynamical and environmental evolution. While a constant SFH with SFR variations of a factor 2-3 over the past few Gyrs was a quite common assumption in early studies of some irregular and dwarf galaxies \citep{marconi95, Grebel20}, recent observations have shown a variety of possible SFHs also for small irregular objects, generally fragile to feedback processes (see \citealt{TolstoyReview09} and references therein).  For dwarf galaxies both within the Local Group and beyond, a fairly moderate star formation activity can be generally assumed averaging over sparse  episodes. In the case of  classical spiral galaxies of the Local group, this assumption should be generally limited to a interval of their global evolution. This also depends on assumptions about quenching efficiency, and a more general Log-normal SFH appears more appropriate when compared with the predictions from numerical simulations.  \citep{Diemer17}. The above assumptions are certainly less appropriate for elliptical galaxies as their old stellar populations are often interpreted in terms of strong initial bursts followed by a long evolution without star formation episodes \citep{Rogers10}, although recent studies suggest more components in their populations \citep{Jegatheesan25}. A common assumption is then not applicable to heterogeneous samples. Semi-analytical models, for example, tend to reduce the parameter space when modeling SFH of heterogeneous samples by assigning simple functional forms to each type \citep{Calura04}. With the above caveats, here we start our investigation with a somehow over-simplified assumption of constant SFH deferring a more appropriate, galaxy type-based  treatment to a future investigation.

\begin{figure*}
\vskip-40pt
\begin{minipage}{0.33\textwidth}
\resizebox{1.\hsize}{!}{\includegraphics{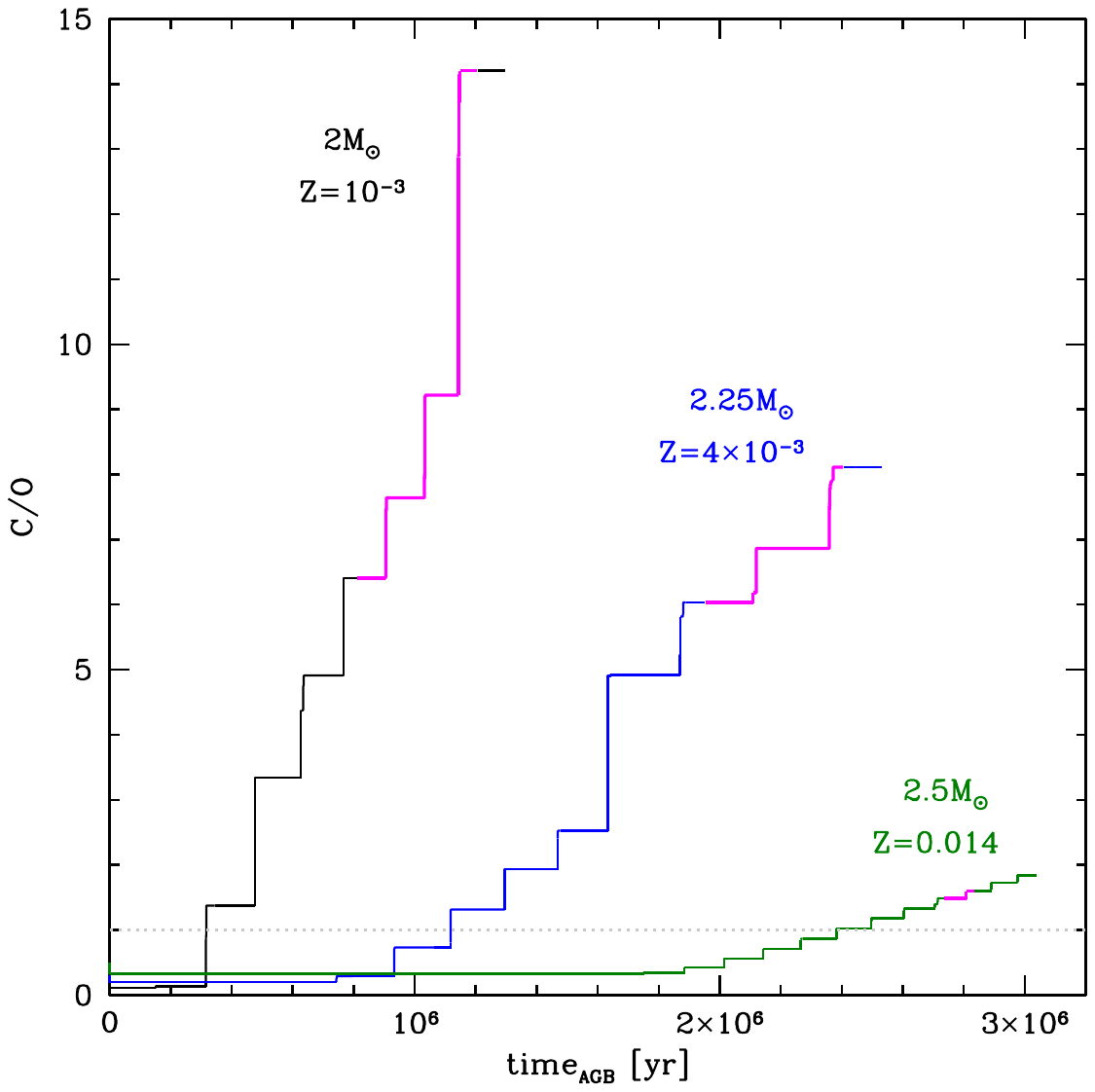}}
\end{minipage}
\begin{minipage}{0.33\textwidth}
\resizebox{1.\hsize}{!}{\includegraphics{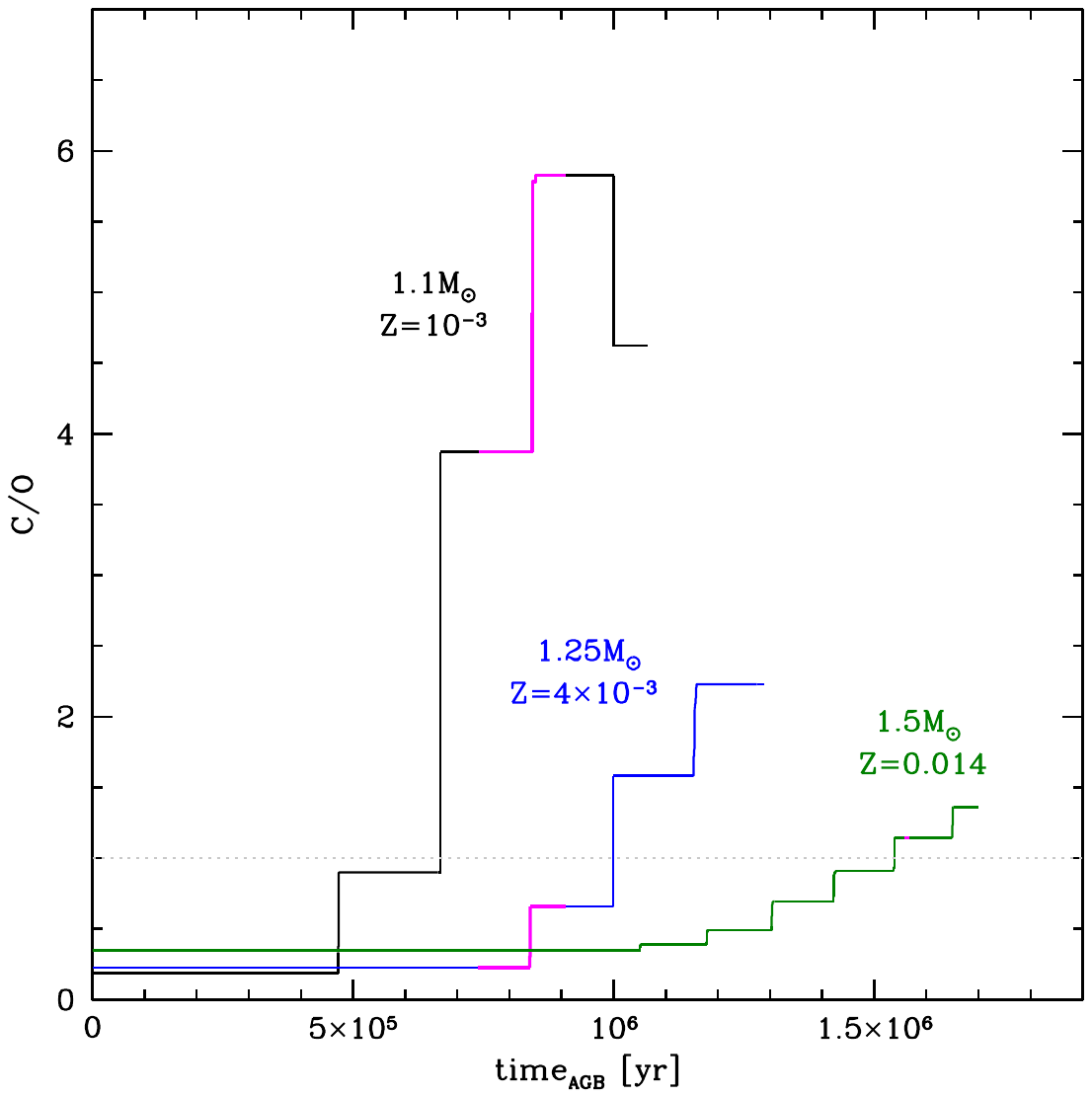}}
\end{minipage}
\begin{minipage}{0.33\textwidth}
\resizebox{1.\hsize}{!}{\includegraphics{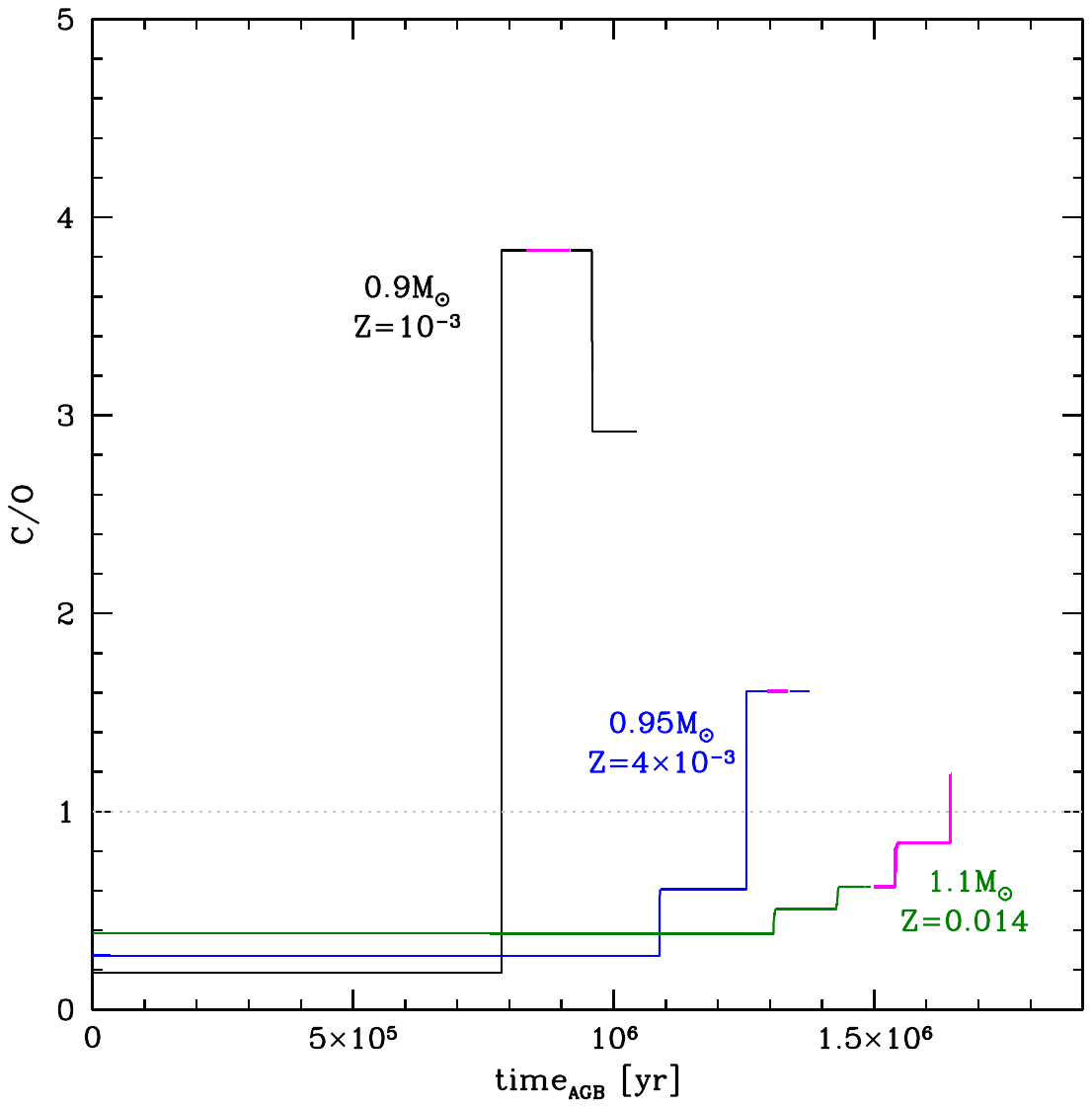}}
\end{minipage}
\vskip-40pt
\caption{Time variation of the surface C/O ratio of 
model stars of metallicity $Z=10^{-3}$ (black lines),
$Z=4 \times 10^{-3}$ (blue) and $Z=0.014$ (green), formed
$\sim 1$ Gyr ago (left panel), $\sim 3$ Gyr ago (middle)
and $\sim 5$ Gyr ago (right). The magenta portion of the lines indicates the phases during which the model stars cross the J region. The grey, horizontal lines
indicate the time when the stars become C-stars. Times are
counted since the beginning of the TP-AGB phase. The masses
reported in the middle and right panels refer to the start
of the core helium-burning phase.
} 
\label{figco}
\end{figure*}

To produce the synthetic populations, we followed the same methodology introduced 
by \citet{flavia14a, flavia15a} and applied
by \citet{cla25, cla26}, which can be summarized into four steps,
briefly discussed below:

\begin{enumerate}

\item{The evolutionary sequences, calculated by means of the ATON code for
stellar evolution \citep{italo, ventura98}, are taken from the Rome library
and were published and discussed in \citet{ventura14}, \citet{ventura18},
\citet{marini21}, \citet{devika23}. For the present investigation we
used stellar models of metallicities $Z=1,2,4,8\times 10^{-3},1.4\times 10^{-2}$, which correspond to the $\rm -1.5 \leq [Fe/H] \leq 0$ range. 
For stars of mass $\rm 2~M_{\odot} < M < 8~M_{\odot}$
the evolutionary sequences were started from the pre-MS phase and evolved 
until the end of the AGB, when almost the entire envelope was lost.
For $\rm M \leq 2~M_{\odot}$ stars the evolutionary sequences were first evolved 
from the pre-MS until the helium flash, then resumed from the core
helium-burning phase, with the same core mass reached at the occurrence of the
helium flash. For the latter sequences started from the core helium-burning an assumption must be made in regard to the amount of mass lost
during their ascent along the RGB, to which we will refer 
as $\rm \delta m_{RGB}$. The results presented here were obtained with 
$\rm \delta m_{RGB}=0.2~M_{\odot}$. This choice is motivated by several 
investigations on the stellar populations of globular clusters, which
demonstrated that RGB mass losses of the order of $\rm 0.2~M_{\odot}$ 
are required to reproduce the various observational features, particularly
the morphology of the horizontal branch \citep{tailo21}. The effects
of different choices of $\rm \delta m_{RGB}$ are discussed in 
Section \ref{mloss}.

The results from stellar evolution modeling allow us to follow the time
variation of the main physical quantities of the star, namely mass,
luminosity, effective temperature, mass-loss rate, and of the surface
chemical composition, during their evolution.
}

\item{For the evolutionary sequences discussed in the previous point,
dust formation in the wind is modeled, following the same approach used 
in previous works on this argument \citep{ventura12, ventura14, ventura18, 
cla25, cla26}. We apply the chemo-dynamical description of the wind proposed by
\citet{fg02, fg06} to some evolutionary stages chosen 
along the AGB (typically, we select 20 points 
during the early-AGB and 10 points between two consecutive thermal pulses 
of the TP-AGB evolution, approximately equally spaced in time). 
The values of the luminosity, effective temperature,
evolving mass and mass-loss rate of the selected evolutionary stages, found as
described in point 1), are used as ingredients to model the
formation and growth of the dust grains, adopting the set of equations
described in detail in \citet{ventura12}. The types of dust species considered
depend on the surface chemical composition: for oxygen-rich 
environments we consider the formation of silicates, alumina dust and
solid iron, while for carbon stars we model the growth of amorphous
carbon, silicon carbide and solid iron. This step leads to the
determination of the overall DPR, of the 
asymptotic size reached by each dust species, of the percentage
contribution of each dust species to the global dust formed in 
the circumstellar envelope, and of the optical depth (similarly to
previous investigations, here we consider the value at the
wavelength value of $10~\mu$m, $\tau_{10}$), for each of the evolutionary 
stages considered.}

\item{The dust mineralogy and the values of $\tau_{10}$ found in point 2
are used to find the synthetic SED for each of the evolutionary stages considered.
This step is accomplished by using the DUSTY code \citep{nenkova99}.
Convolving the synthetic SED with the transmission curves
of the observational filters enables the computation of the
evolutionary tracks for the model stars of different mass and
metallicity in the various CMDs of interest.
In the present investigation, as discussed in the previous section, we focus on 
the $\rm (J-K_s, J)$ CMD.}

\item{The final step consists of the construction of a synthetic
distribution of stars in the CMD, using a population 
synthesis approach, as described in \citet{cla25}. Briefly,
we consider the time evolution since 14 Gyr ago to the present,
assuming, as discussed above, that the SFR has been constant
across the various epochs, and that the initial mass function
(IMF) is described by a \citet{kroupa01} law. The timescale
of metal enrichment, $\rm \tau_{Z}$, is used as a free parameter.
The distribution of stars on the CMD is found by considering
the time evolution of stars of different mass and metallicity,
and then by interpolation among the different tracks, 
calculated as described in point 3) above.}
\end{enumerate}
For the construction of the synthetic populations discussed in this section, we adopted a SFH characterized by a SFR of 0.3 $\rm M_{\odot}/{yr}$ extended over the entire life of the galaxy (14 Gyr). This choice results in a total stellar mass of approximately $\rm 4.2 \times 10^9 \, M_{\odot}$, which ensures that the JAGB region is populated by a statistically robust sample of artificial stars. It is important to emphasize that the resulting morphology of the JLFs, as well as the determination of key parameters like the mean and the peak J magnitudes, are numerically independent of the specific choice of the SFR value. From an astrophysical standpoint, the relative shape of the JLF is fundamentally determined by the product of the IMF and the evolutionary crossing times ($\rm \tau_{JAGB}$) of stars within the $\rm 1-3 \, M_{\odot}$ range. Since the SFR acts primarily as a global normalization factor, changing its value scales the total number of stars but does not alter their relative distribution across magnitude bins, provided the total count is high enough to minimize stochastic Poissonian fluctuations.

\begin{figure*}
\vskip-40pt
\begin{minipage}{0.33\textwidth}
\resizebox{1.\hsize}{!}{\includegraphics{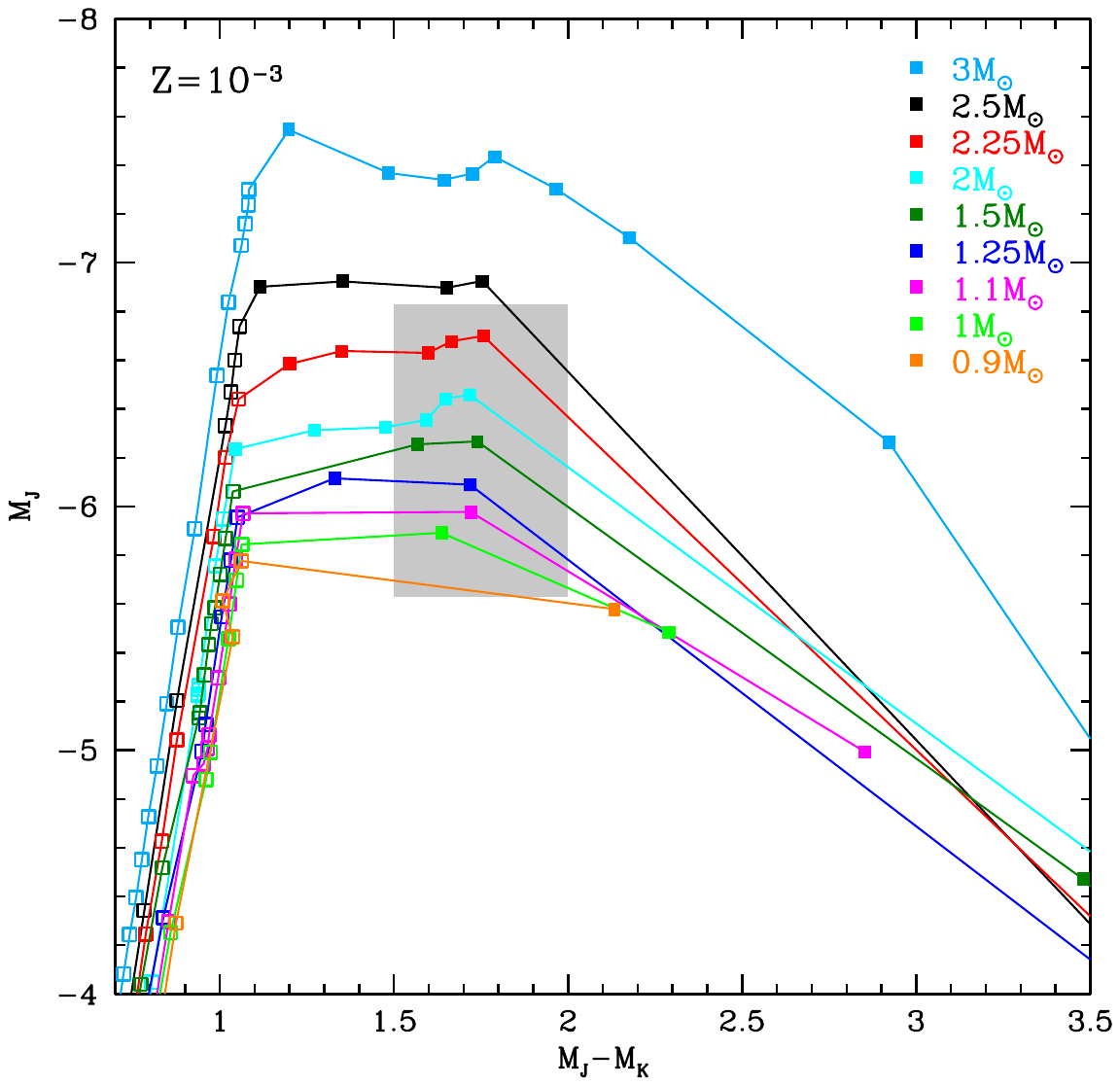}}
\end{minipage}
\begin{minipage}{0.33\textwidth}
\resizebox{1.\hsize}{!}{\includegraphics{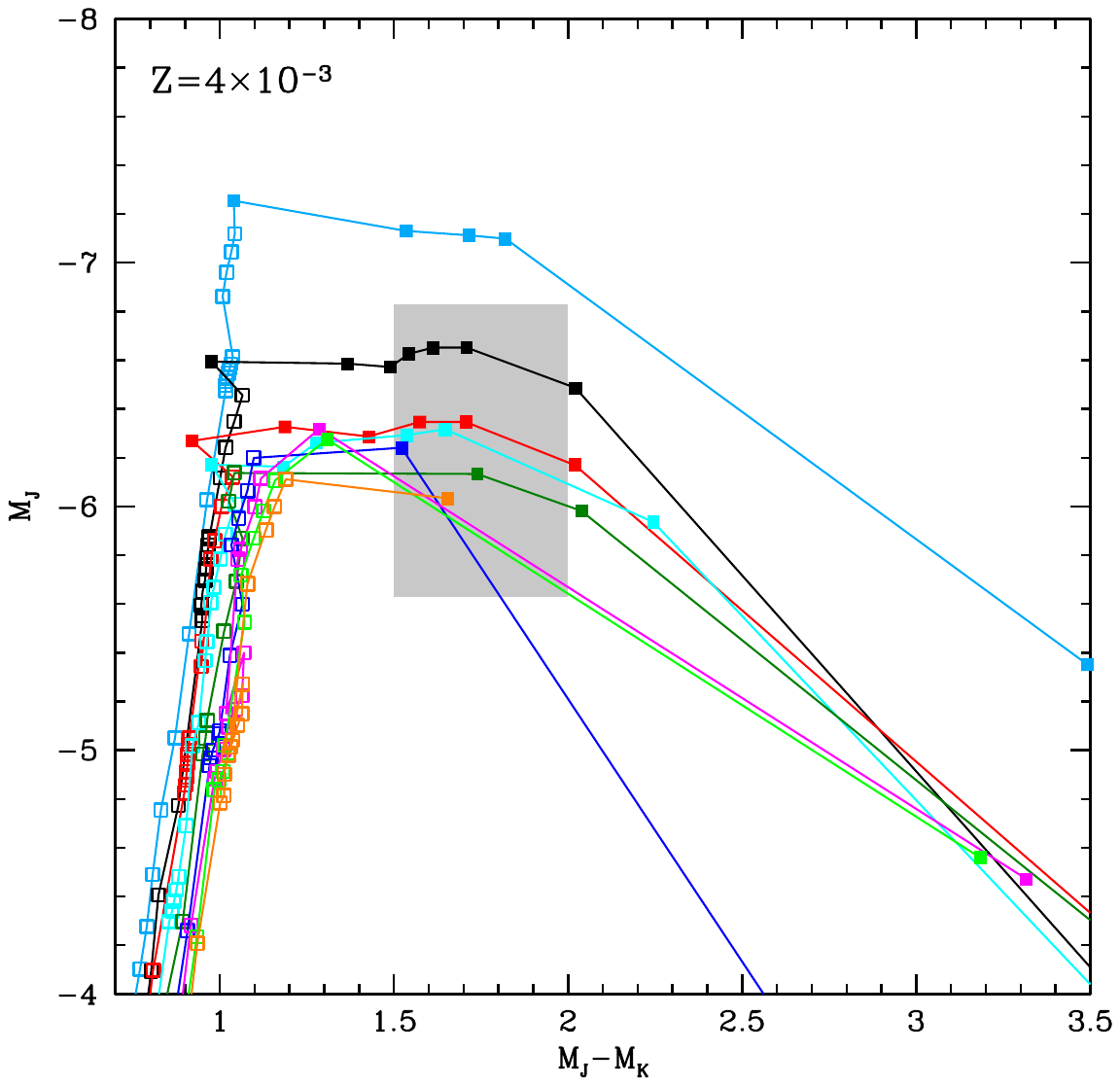}}
\end{minipage}
\begin{minipage}{0.33\textwidth}
\resizebox{1.\hsize}{!}{\includegraphics{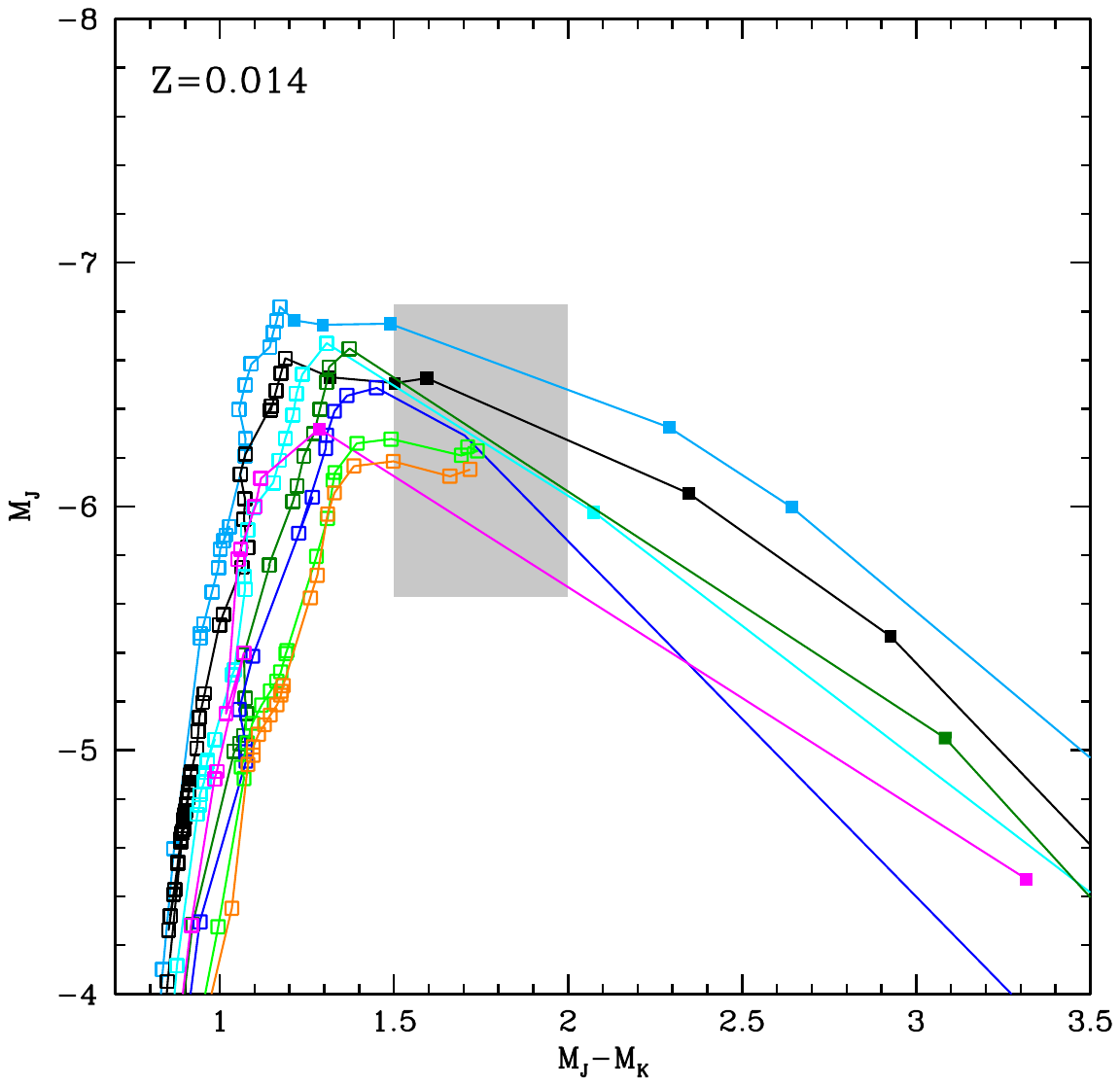}}
\end{minipage}
\vskip-40pt
\caption{Evolutionary tracks of stars of different mass and
metallicity on the color-magnitude $\rm (J-K_S, J)$ diagram.
For $\rm M\leq 2~M_{\odot}$ the masses are taken at the
beginning of the core helium-burning phase. The points along
each track refer to the phases of maximum luminosity experienced 
during an inter-pulse phase. Open symbols refer to the phases during which the star is O-rich, whereas full points indicate C-rich phases. The grey-shaded box indicate the
JAGB box defined by \citet{magnus24}.
} 
\label{figcmd}
\end{figure*} 

\section{The role of mass and metallicity in the carbon 
enrichment and infrared colors of AGB stars}
\label{tracks}

\citet{ventura22} divided AGB stars into three groups, distinguished on
the basis of the modification of the surface chemical composition. 
Group I consists of low-mass stars, which, after only 
a few third dredge-up (TDU) events, lose the external envelope when the surface $\rm C/O$
is still below unity. This behavior is common to the stars that start 
the core helium-burning phase with a mass below a threshold value $\rm M_{carb}$,
of the order of $\rm 1~M_{\odot}$, which increases with the 
metallicity \citep{devika23}: this is because metal-poor stars become 
C-stars more easily, owing to the low initial oxygen content.
Group II includes stars of mass between $\rm M_{carb}$ and 
an upper threshold $\rm  M_{HBB} \sim 3-4~M_{\odot}$, which become carbon 
stars during the AGB phase. In these stars the surface carbon gradually
increases 
during the entire AGB lifetime, owing to repeated TDU events experienced, 
until the envelope is lost. Group III is composed of massive AGB stars,
of initial mass above $\rm M_{HBB}$. These stars experience hot bottom 
burning (HBB) \citep{sack, blo91}, a process that destroys carbon nuclei 
in the outer layers of the star via proton-capture nucleosynthesis at the 
base of the convective envelope. Consequently, these stars do not become
C-stars. $\rm M_{HBB}$ decreases with metallicity, as HBB is activated 
more easily in metal-poor stars \citep{flavia18}. 

The stars belonging to the three groups populate 
different regions of the CMD: this is due partly to the
obvious differences in the luminosities, which increase with the 
progenitor's mass, and partly to the fact that the path
followed by the evolutionary tracks 
is highly sensitive to whether and when the stars reach the C-star stage.
Indeed, the achievement of the $\rm C/O>1$ condition is 
accompanied by a significant increase in the surface molecular
opacities \citep{marigo02}, which triggers the
expansion and cooling of the surface regions, and
consequently the increase in the mass-loss rate and in
the density of the wind. These conditions turn extremely 
favourable to the formation of carbon dust, which effectively
starts when the carbon excess with respect to oxygen exceeds
$\sim 10\%$, so that there is a sufficiently large number
of carbon molecules available to condense into carbon 
grains\footnote{The achievement of the $\rm C/O>1$ condition
is per se not sufficient to the formation of carbon dust, 
because the CO molecules are very stable, thus only the
carbon in excess to oxygen is available to condense into
solid particles \citep{fg06}.}. 
The formation of carbon dust makes the SED 
to shift to the NIR and MIR spectral regions, so
the stars evolve to the red side of the CMD.

Since most, if not all, the stars in the J region are
carbon stars, in the present analysis we focus our attention 
on the stars belonging to group II. Indeed the stars in group III
are of little interest for the comprehension of the stellar population
of the J region, because they are too bright to evolve inside the box suggested by 
\citet{magnus24} to define the J region, and also their statistical 
weight would be negligible 
even if a different box was chosen, owing to their short evolutionary 
timescales. Even the stars in group I, which never become C-stars, rarely
enter the J region due to their low dust production, with the sole exception of the solar
metallicity case, which will be discussed in the following sections.

\subsection{The start and the timescale of the surface carbon enrichment}
\label{csuo}
As the formation of carbonaceous dust begins when there is a sufficiently
large excess of carbon with respect to oxygen in the surface regions of the star,
the key parameter determining whether a star enters
the J region during its evolution is the
surface $\rm C/O$ ratio. The variation of this quantity during the AGB phase for stars of different mass and chemical composition is
shown in Fig.~\ref{figco}. The stars of different metallicity
evolving on timescales of 1, 3, 5 Gyrs are shown in the left,
middle, and right panels, respectively. This comparison, aimed at identifying when stars of different ages and metallicities become C-stars, will prove 
extremely important for understanding the contribution of stars 
formed during the various epochs to the population of the
J region. Fig.~\ref{figco} is limited to the TP-AGB phase, and the
time on the abscissa is measured from the beginning of the TP-AGB phase.

For stars formed 1 Gyr ago, we consider model stars of mass
$\rm 2~M_{\odot}$, $\rm 2.25~M_{\odot}$ and $\rm 2.5~M_{\odot}$, for
the $\rm Z= 10^{-3}$, $\rm Z=4 \times 10^{-3}$ and $\rm Z=0.014$
cases, respectively. These are the stars accumulating the
largest amounts of surface carbon, whose mass fraction eventually exceeds 
$1 \%$. \citet{flavia15a} associated with the late evolutionary phases
of these objects the reddest sources populating the color-color
$([3.6]-[4.5], [5.8]-[8.0])$ diagram of the LMC, and concluded that
they provide the most significant contribution to the overall DPR
of the galaxy. Their peculiarity arises from
the fact that they reach the C-star stage after only
a small fraction of the envelope mass is lost. Consequently, they experience several TPs after becoming C-stars, allowing the surface carbon abundance - and therefore the DPR - to increase considerably. The fraction of the AGB lifetime 
spent as a C-star decreases from $\sim 80\%$, for the $\rm Z= 10^{-3}$
model star, to $\sim 15\%$ in the solar case. This is because higher-metallicity
stars form with a larger content of oxygen, which hinders the
achievement of the C-star stage. 

As regards the stars formed 3 Gyr ago, the model stars reported in the
middle panel of Fig.~\ref{figco} are characterized by masses at the start of
the core helium-burning phase of $\rm 1.1~M_{\odot}$, $\rm 1.25~M_{\odot}$ and 
$\rm 1.5~M_{\odot}$, for the $\rm Z= 10^{-3}$, $\rm Z=4 \times 10^{-3}$ and 
$\rm Z=0.014$ cases, respectively. The most relevant difference with respect
to the higher-mass counterparts, discussed above, is that
these stars become C-stars only during the final part of the AGB phase, thus
they experience only a few TDU events after the surface C/O ratio exceeds unity in the surface regions. Consequently, the fraction of the AGB lifetime 
spent as C-stars is significantly shorter than the 1 Gyr old counterparts,
and the final $\rm C/O$ ratios are smaller. In the specific case of the solar metallicity model
star, the C-star phase is limited to the last two inter-pulse phases, 
and the fraction of time spent as a C-star is found to be below $10\%$.

Finally, in the case of the low-mass stars formed 5 Gyr ago, the effect of metallicity becomes even more important than in the cases discussed above.
Indeed only the metal-poor model star (with an initial mass of $\rm 0.9~M_{\odot}$ at the start
of the core helium-burning phase) spends a non
negligible fraction ($\sim 25\%$) of the AGB lifetime as a C-star, while 
the $\rm 0.95~M_{\odot}$ model star of $\rm Z=4\times 10^{-3}$ become a C-star only 
when the last TP takes place, and the solar metallicity model star of mass
$\rm 1.1~M_{\odot}$ fails to reach the C-star stage.

\subsection{The evolutionary tracks on the $\rm (J-K_s, J)$ diagram}
\label{tracce}
As discussed earlier in the present section, the change in the 
surface chemical composition, particularly in the $\rm C/O$ ratio,
affects the morphology of the evolutionary tracks on the CMD.
Some selected tracks of stars of different mass and 
metallicity are shown in Fig.~\ref{figcmd}. Each point along the tracks 
corresponds to a specific stage during the inter-pulse phase, when
the luminosity attains the maximum value: thus, the spacing between the points indicates how fast the tracks move across the CMD.

In general, the evolutionary tracks move vertically 
during the first part of the
AGB evolution, during which they are practically dust-free, and 
the J flux increases as a consequence of the gradual rise in the
luminosity. When carbon dust formation begins, the tracks
move to the red, owing to the gradual shift of the SED to the 
NIR. The J flux increases until $\rm (J-K) \sim 1.7$ mag,
then decreases, because the peak of the SED moves to wavelengths
$\rm \lambda > 1~\mu m$, as carbon accumulates in the surface layers of the star, and the DPR
increases. The only exceptions to this common
behavior are the $\rm 3~M_{\odot}$ model stars of metallicity
$Z=10^{-3},4 \times 10^{-3}$, and the $\rm M \leq 1~M_{\odot}$ model
stars of solar chemical composition: in these cases the redward excursion of the evolutionary tracks is associated with the formation 
of silicates. The theoretical necessity of including dust formation to explain the NIR colors of the JAGB population is strongly corroborated by observational evidence. If we consider, for example, the spectroscopically confirmed carbon stars (Wood et al. (2011)) populating the J region, they show dust features (e.g. silicon carbide emission) in their spectra.

\begin{figure*}
\vskip-40pt
\begin{minipage}{0.48\textwidth}
\resizebox{1.\hsize}{!}{\includegraphics{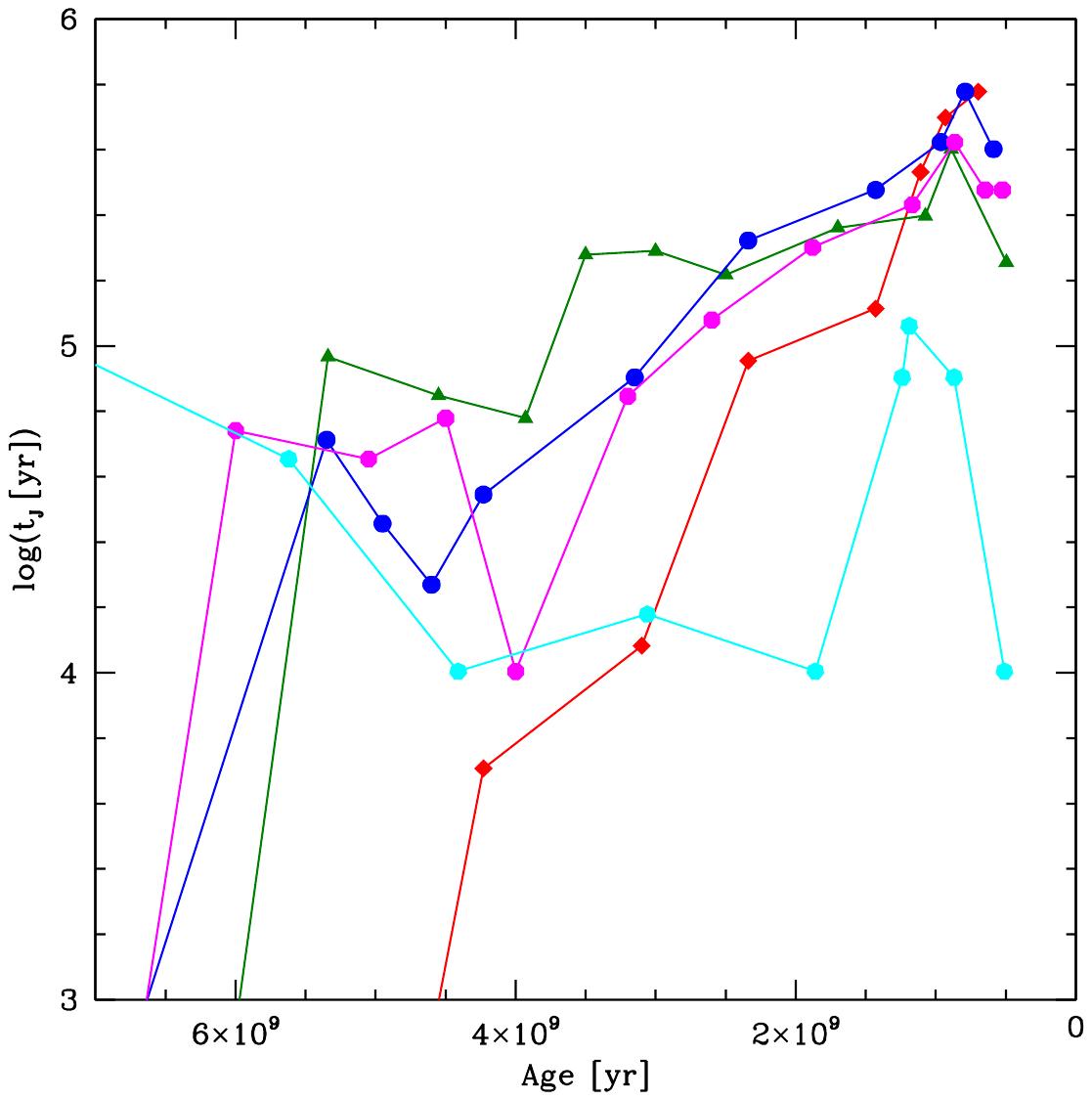}}
\end{minipage}
\begin{minipage}{0.48\textwidth}
\resizebox{1.\hsize}{!}{\includegraphics{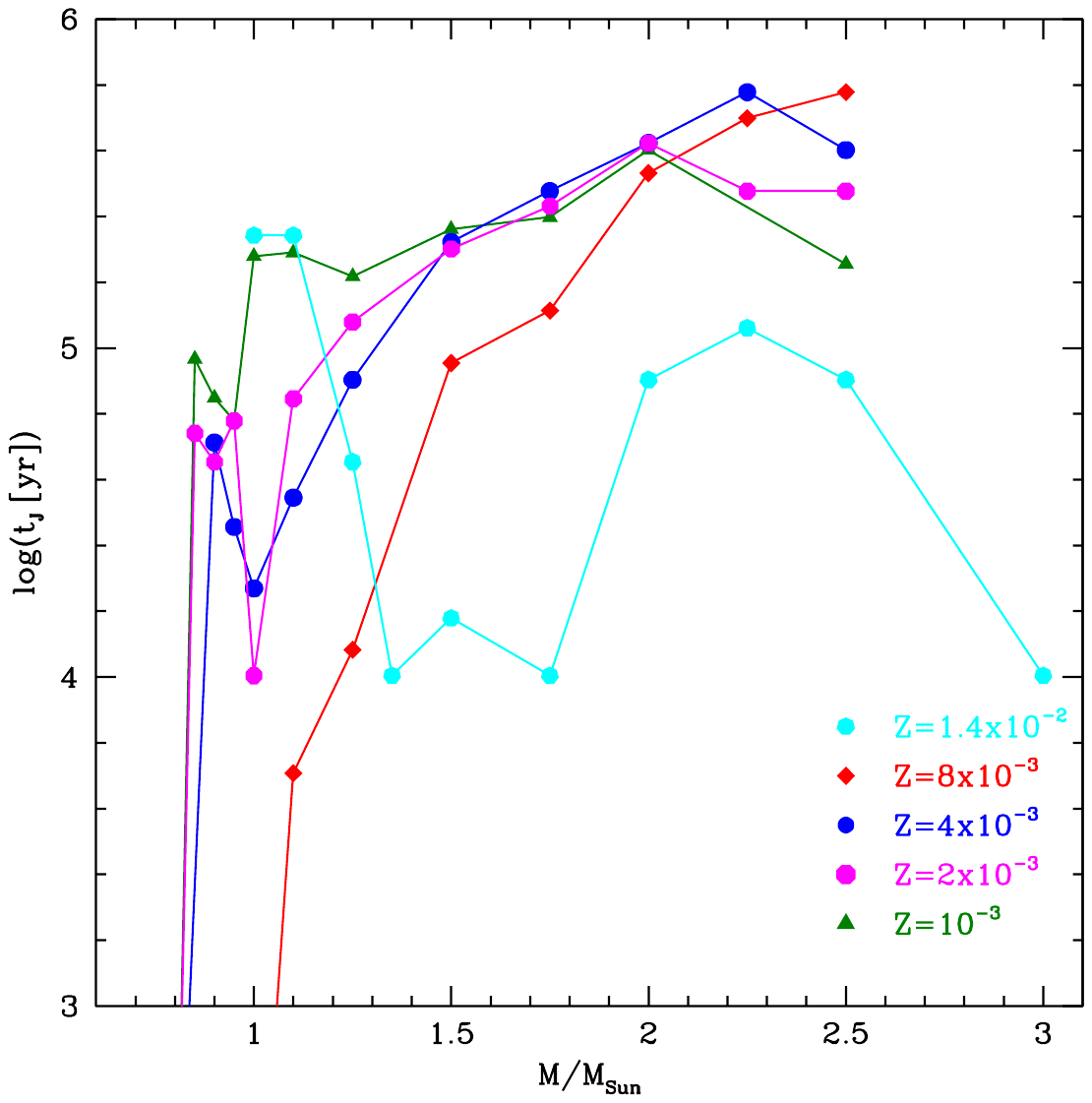}}
\end{minipage}
\vskip-40pt
\caption{Times spent within the J region by model stars
of different mass and metallicity as a function of the
progenitor's mass (left panel) and of the formation epoch
(right). The masses for $\rm M \leq 2~M_{\odot}$ stars refer to the
start of the core helium-burning phase. For masses below
$\rm 1.5~M_{\odot}$ we consider an average mass loss during the
RGB evolution of $\rm 0.2~M_{\odot}$, thus the formation epochs
reported on the abscissa of the right panel refer to those
of the model stars of mass $\rm 0.2~M_{\odot}$ higher than 
those reported on the abscissa of the left panel. Each sequence is limited to the masses whose the evolutionary tracks enter the J region.
} 
\label{figtimes}
\end{figure*}

The metallicity effects can be deduced
by comparing the morphology of the evolutionary tracks of
low-mass stars of different $Z$. For the reasons given before, 
metal-poor stars become C-stars after experiencing only a few
TPs, when the core mass, hence the luminosity, is lower than 
that of the higher-metallicity counterparts. This reflects
into the J flux at which the transition to the red side 
of the CMD takes place, which is smaller the lower the
metallicity.

\subsection{The crossing time of the J region}
\label{cross}
Characterizing the stellar population of the J region
requires knowledge of the time that the stars spend in that 
part of the CMD. As discussed earlier in this section, the stars
evolve within this region of the CMD during 
an intermediate evolutionary phase, which starts soon after the C-star stage
is achieved and lasts until the DPR attains
values so large that the $\rm (J-K_s)$ colors increase, the
J flux decreases, and the stars will no longer populate
the JAGB box. 

Fig.~\ref{figtimes} shows the times $\rm \tau_{JAGB}$ spent in the J
region of the CMD by model stars of different mass and metallicity,
as a function of the initial mass (left panel) and of the formation
epoch (right). As discussed in \citet{cla26}, the stars spending the 
longest time within the JAGB box, slightly below 1 Myr, are the progeny of 
$\rm 2-2.5~M_{\odot}$ stars, formed $\sim 1$ Gyr ago. This is 
consistent with the discussion
of Section \ref{csuo} and is related to the fact that these objects 
reach the C-star stage when only a small fraction of the envelope is lost, 
thus the increase in the surface carbon is gradual and the transition to
the very red side of the CMD is delayed. This can be seen also 
in Fig.~\ref{figcmd}, where we note that the evolutionary tracks of 
$\rm 2-2.5~M_{\odot}$ stars remain within
the JAGB box for several inter-pulse phases. 

As far as 
sub-solar metallicities are concerned, the $\rm \tau_{JAGB}$ 
vs mass trend is positive, given the small fraction of time 
during which low-mass stars evolve as C-stars, which can be 
deduced by comparing the results shown in the middle and 
right panels of Fig.~\ref{figco} with those reported on the 
left panel of the same figure.
We note in Fig.~\ref{figcmd} that the evolutionary tracks of 
low-mass stars stay within the JAGB box for at most a couple
of inter-pulse phases, before moving toward the red, low luminosity
side of the plane. Solar metallicity model stars are an exception to this general 
behavior, because after becoming C-stars they cross
the J region very fast, owing to the rapid expansion and cooling
of the outermost regions. Unlike the lower-metallicity cases, the stars
spending the longest time within the J region ($\rm \sim 2\times 10^5$ yr) 
are the progeny of stars with masses $\rm 1-1.2~M_{\odot}$: these stars
evolve into the J region during the final AGB phases, when they
are oxygen-rich, and the redwards excursion of the evolutionary
tracks is due to the formation of silicates in the circumstellar
envelope, in turn favored by the cool surface temperatures and by
the large amount of gaseous silicon.

The slope of the $\rm \tau_{JAGB}$ vs mass (or formation epoch) relation
reported in Fig.~\ref{figtimes} is significantly affected by the 
metallicity: this trend, discussed in \citet{cla26}, is due to
the fact that metal-rich, low-mass stars reach the C-star stage only 
after a significant fraction of the envelope is lost, and thus evolve
as C-stars only during the very final AGB phases. This can be deduced
by comparing the fraction of time spent in the C-star phase by the
model stars of different metallicity, in the middle and right 
panels of Fig.~\ref{figco}. 

The results presented so far indicate that the vast majority, if not all,
the stars populating the J region, formed between 1 and 6 Gyr ago. 
This is the time interval of interest for a critical evaluation of
the various factors entering the characterization of the stars in the
JAGB box and affecting their luminosity distribution. The star
formation that occurred around 1 Gyr ago is crucial to determine the
numerical consistency of the sample of stars within the J region,
because that is the formation epoch of the stars that spend more time in that region than any other stars, namely
the progeny of $\rm 2-2.5~M_{\odot}$ stars. This is even more true
in metal-rich environments, because low-mass stars barely become
C-stars, thus fail to enter the J region. We expect that the
JAGB population of the galaxies in which stars of solar metallicity formed
earlier than $1-2$ Gyr ago is significantly different from that
of the other galaxies, owing to the peculiar behavior of 
solar metallicity stars.

\begin{figure*}
\vskip-40pt
\begin{minipage}{0.48\textwidth}
\resizebox{1.\hsize}{!}{\includegraphics{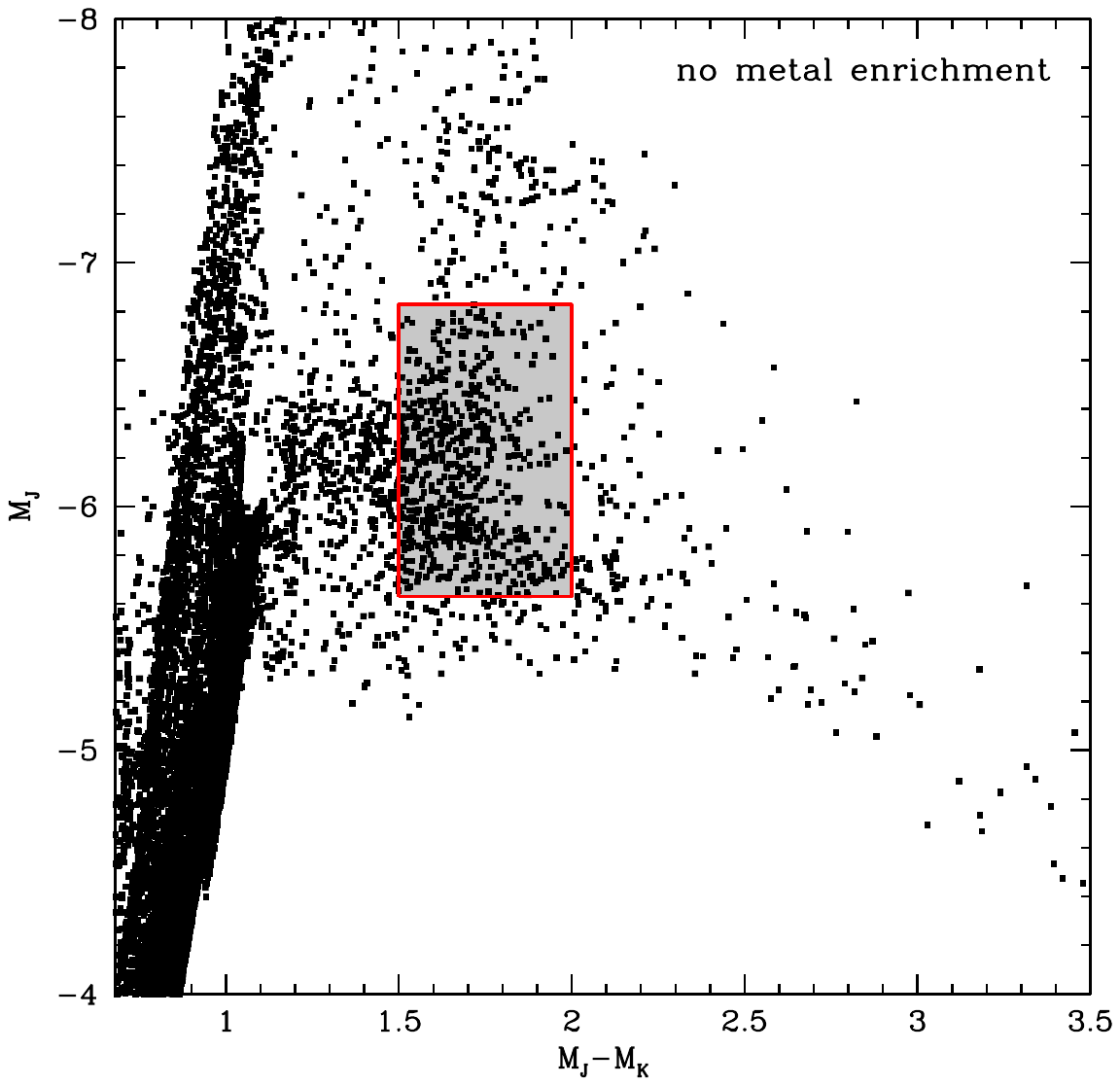}}
\end{minipage}
\begin{minipage}{0.48\textwidth}
\resizebox{1.\hsize}{!}{\includegraphics{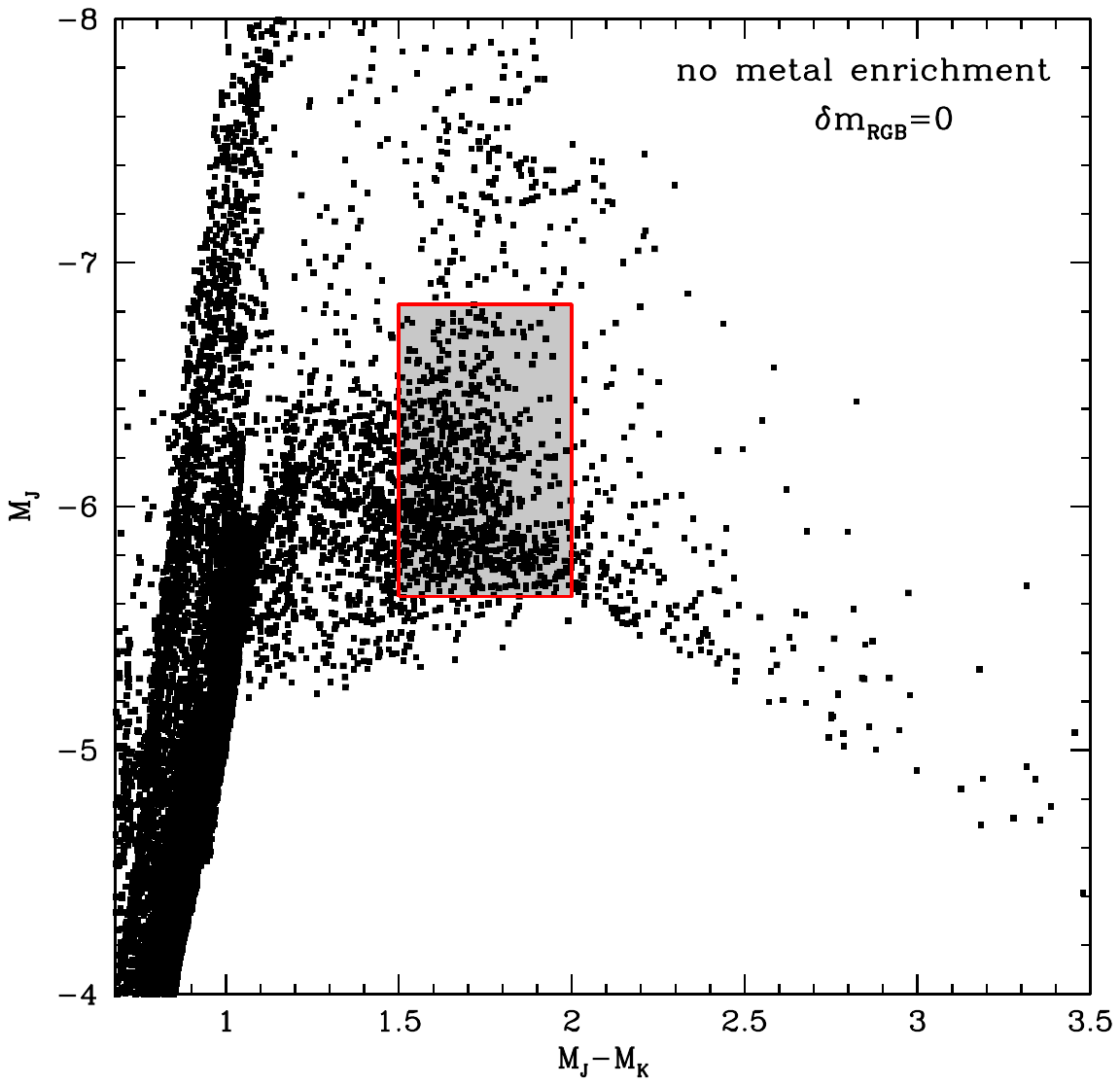}}
\end{minipage}
\vskip-90pt
\begin{minipage}{0.48\textwidth}
\resizebox{1.\hsize}{!}{\includegraphics{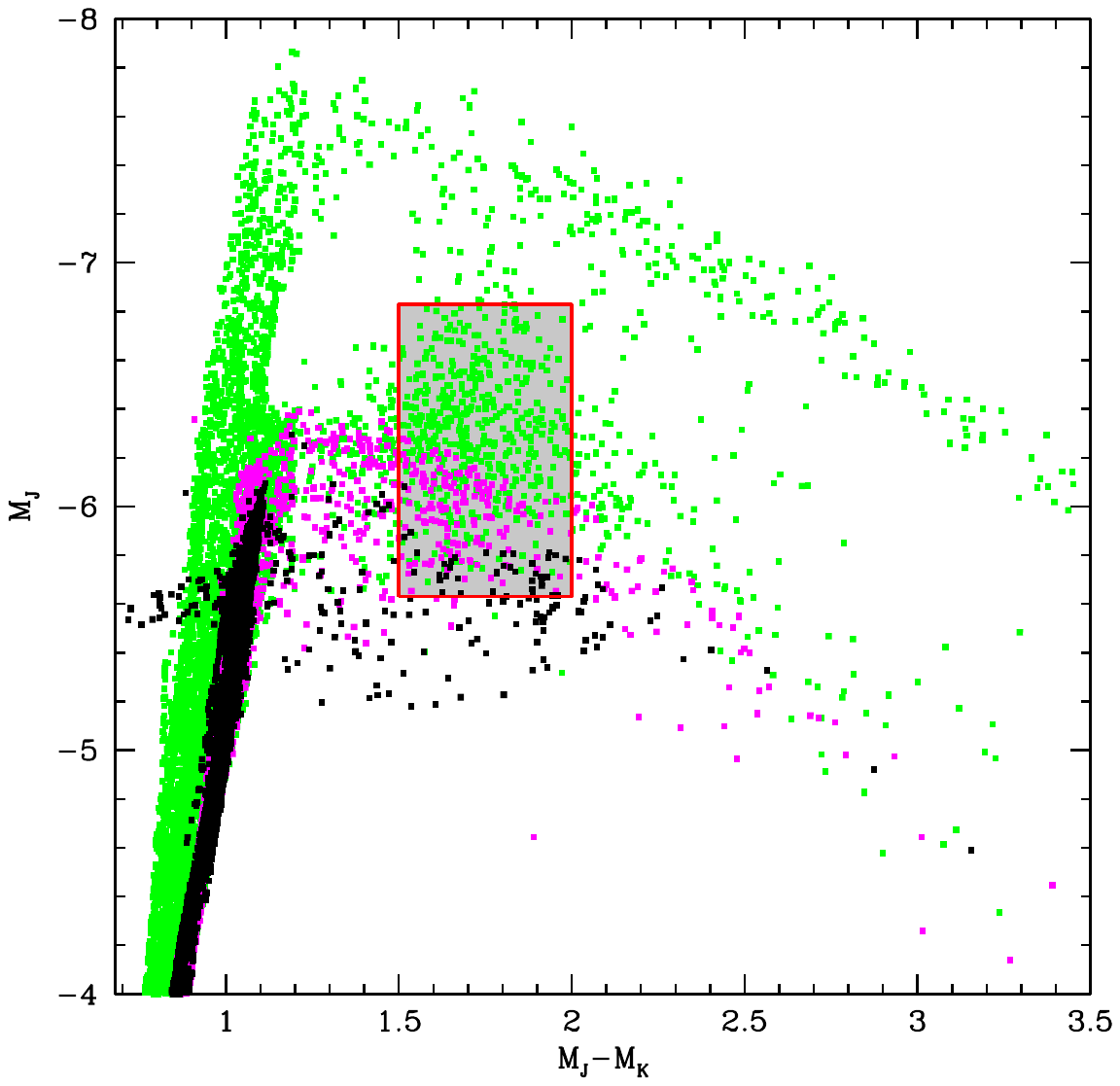}}
\end{minipage}
\begin{minipage}{0.48\textwidth}
\resizebox{1.\hsize}{!}{\includegraphics{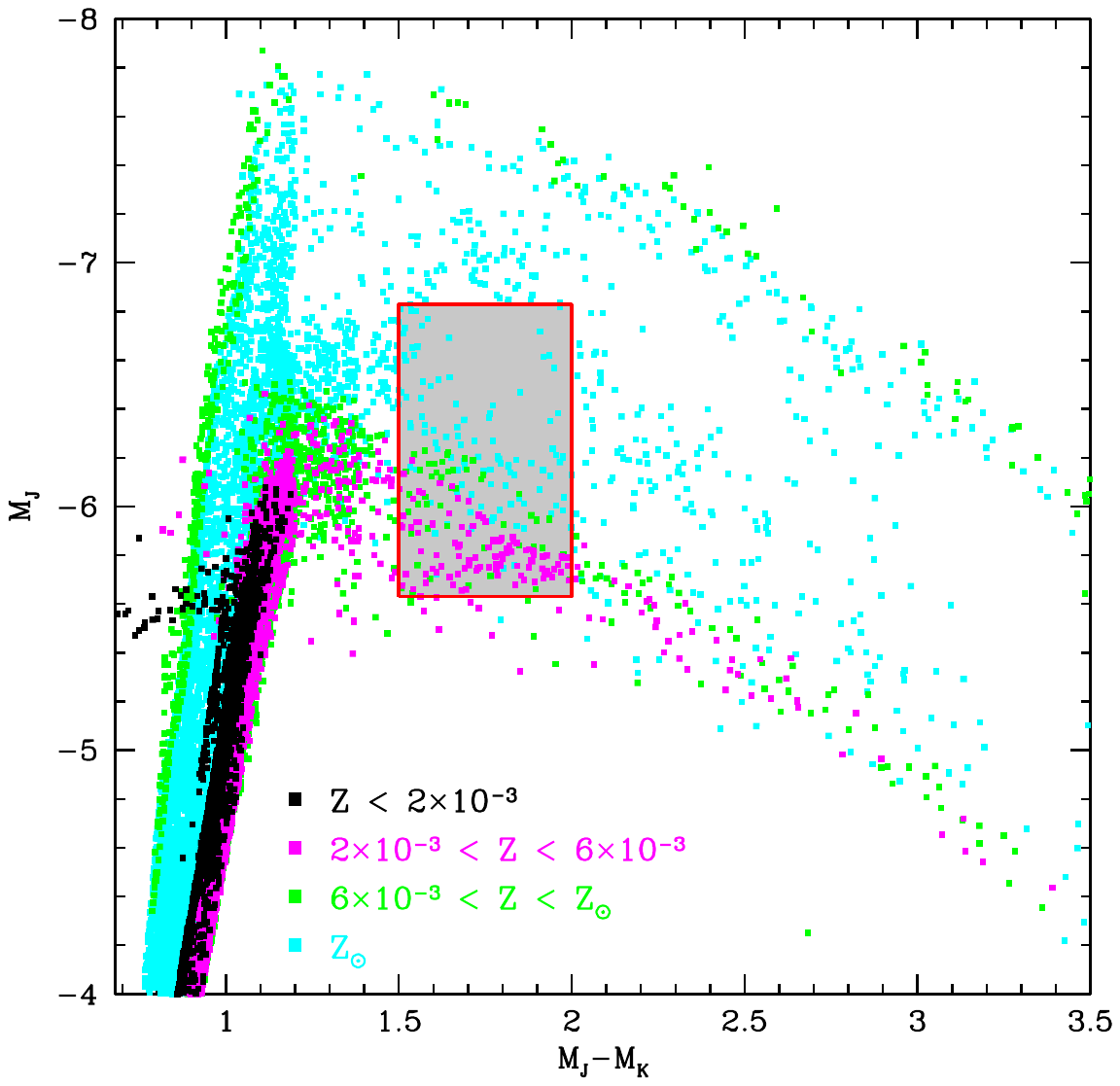}}
\end{minipage}
\vskip-60pt
\caption{Synthetic distributions of stars in the $\rm (J-K_s, J)$
diagram, obtained by assuming constant SFR and
different metal enrichments. The different panels correspond to the
cases with no metal enrichment (top, left), with the metallicity
gradually increasing until $Z=8\times 10^{-3}$ (bottom, left),
and $Z_{\odot}$ (bottom, right). The top right panel reports
the results obtained with no metal enrichment and no mass loss
during the RGB phase. The different colours refer to stars with
different metallicity. The grey-shaded region corresponds to the
JAGB box proposed by \citet{magnus24}.
} 
\label{figmatteo}
\end{figure*} 

\section{The J luminosity function of the stars in the J region}
\label{jlf}

\begin{figure}
\begin{minipage}{0.48\textwidth}
\resizebox{1.\hsize}{!}{\includegraphics{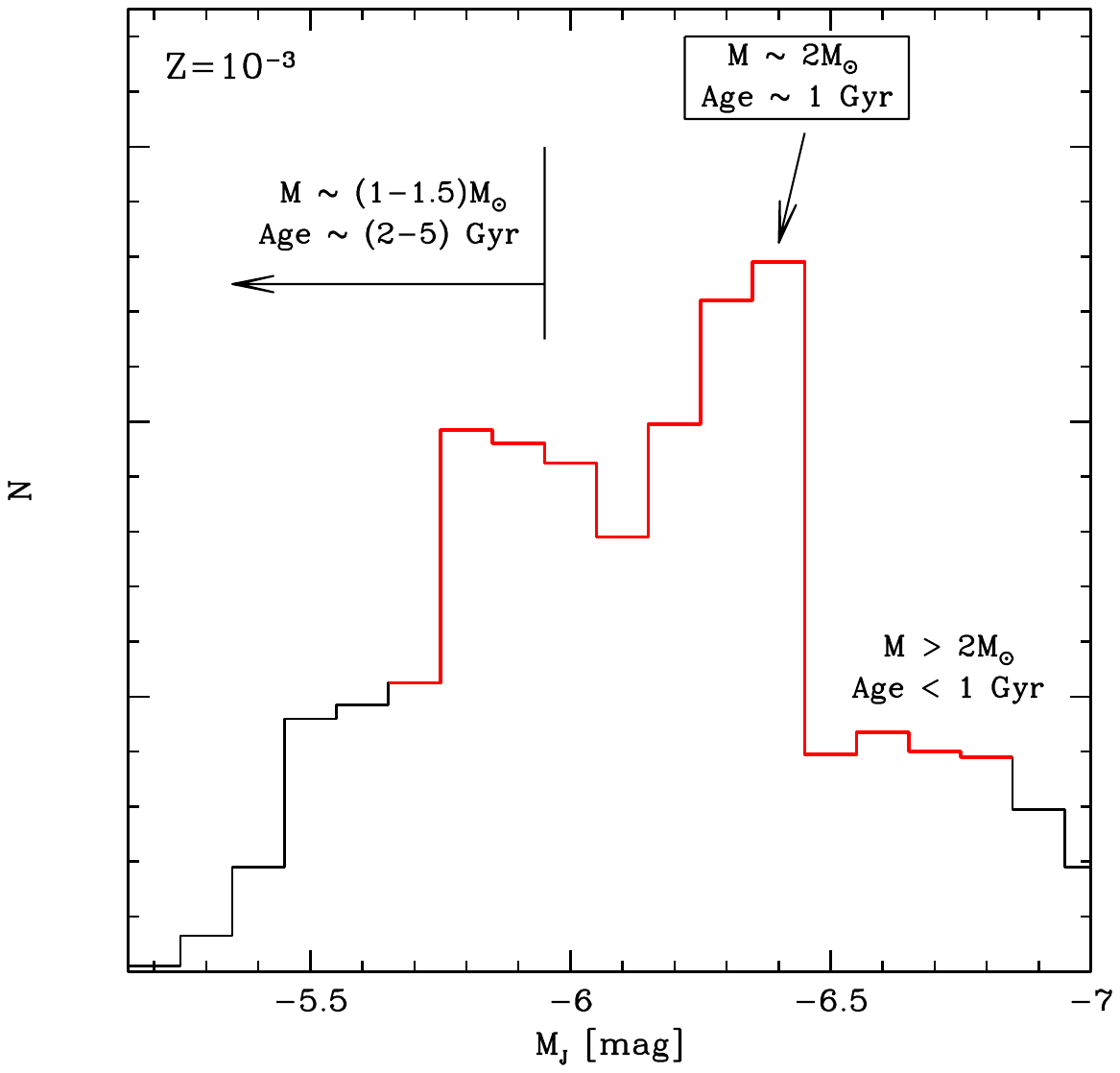}}
\end{minipage}
\vskip-60pt
\caption{The JLF of the stars populating the
$\rm 1.5 < (J-K_S) < 2$ mag region of the color-magnitude
$\rm (J-K_S, J)$ diagram, in a galaxy characterized by
constant SFR, with metallicity $Z=10^{-3}$. The thick, red part
of the histogram refers to the stars in the box suggested by
\citet{magnus24}, with $\rm -6.83 < M_J < -5.63$ mag.
}
\label{fz1m3}
\end{figure}

The arguments presented in the previous section provide the key elements for understanding how stars evolve into the J region, and thus how the expected JLF changes
according to the SFH of galaxies.
As discussed in Section \ref{intro} and Section \ref{input}, in the present
analysis we apply the population synthesis approach
described in Section \ref{input} to galaxies where the SFR was constant
and the metal enrichment of the interstellar medium proceeded with
different timescales. We start from the basic case, that no metal
enrichment occurred, and end up with the case that the metal 
enrichment eventually led to the formation of stars with solar chemical 
composition. A few examples of the synthetic CMDs obtained are
shown in Fig.~\ref{figmatteo}.

\subsection{The case of no metal enrichment}
\label{z1m3}
The first case considered is a galaxy where the metallicity
of the stars formed did not change with time and was constant
at $Z=10^{-3}$. The selection of $Z=10^{-3}$ as the baseline metallicity for our simulations is primarily motivated by the fact that this value represents a realistic floor for the most metal-poor components typically found in the dwarf and irregular galaxies investigated \citep{marconi95, Grebel20, TolstoyReview09}. By focusing on the range $10^{-3} < \rm Z <1.4 \times 10^{-2}$, we effectively encompass the metallicity spread observed in most Local Group galaxies while providing a robust framework to evaluate the impact of chemical enrichment rates. Furthermore, the adoption of a constant SFR coupled with a "no-metal-enrichment" scenario provides a realistic approximation for many dwarf and irregular galaxies both within and outside the Local Group \citep{marconi95, Grebel20, TolstoyReview09}. These systems often exhibit moderate and persistent star formation activity when averaged over long timescales, resulting in a very slow chemical enrichment of the interstellar medium. For such galaxies, the change in metallicity during the critical 1–6 Gyr window is effectively negligible. This case also acts as a fundamental benchmark to isolate the impact of the SFH itself on the JLF, allowing us to assess the method's intrinsic stability before introducing the complexities of rapid chemical evolution.

According
to the results regarding the $Z=10^{-3}$ metallicity reported in 
Fig.~\ref{figtimes}
(see the dark green line), we deduce that the stars currently evolving
in the J region formed between $\sim 1$ Gyr and $\sim 6$ Gyr ago.
The mass (and thus age) distribution of the stars is determined by:
1) whether the stars of a given mass enter the J region;
2) the time spent in the J region; 3) the duration of the epochs 
during which the stars entering the J region formed; 4) the IMF
of the stars formed. In the present case, point (1)
restricts the attention to the $\rm 1-2.5~M_{\odot}$ mass range.
Point (2) favors the contribution of stars descending from $\rm 2-2.5~M_{\odot}$ 
progenitors, as shown in Fig.~\ref{figtimes}. The formation
of low-mass stars is favored on the basis of points (3) and (4), 
given the steepening of the evolutionary time-mass relation, towards 
the low-mass domain. Overall, in the present case we find that 
$\sim 35\%$ of the J population descend from
$\rm 1-1.3~M_{\odot}$ progenitors, $\sim 30\%$ from $\rm 1.7-2~M_{\odot}$ stars,
the remaining objects being distributed between $\rm 1.3-1.7~M_{\odot}$
and $\rm M > 2~M_{\odot}$ stars.

The synthetic distribution of the stars obtained in this case is
shown in the top, left panel of Fig.~\ref{figmatteo}, while 
Fig.~\ref{fz1m3} shows the JLF obtained. The region of the CMD considered 
is that within the color strip $\rm 1.5 < J-K_S < 2$ mag. The plot
in Fig.~\ref{fz1m3} refers to the general LF, with no restrictions on 
the J flux of the stars; the thick portion of the LF corresponds to the 
JLF of the stars falling within the box defined by \citet{magnus24}, 
where the color constraint is complemented by the requirement that 
$\rm -6.83 < M_J < -5.63$ mag.

The LF exhibits a prominent peak at $\rm M_J = -6.4$ mag, which is due
to the stars descending from progenitors of the masses clustered
around $\rm 2~M_{\odot}$, which formed around 1 Gyr ago. The presence
of this peak is consistent with the results shown in Fig.~\ref{figtimes},
indicating that these are the stars that evolve longest 
within the J region of the CMD. A further, more subtle reason for the 
presence of the primary peak in the JLF is that $\rm \sim 2~M_{\odot}$
is the upper mass limit for the occurrence of the helium flash at the
metallicity $Z=10^{-3}$, and that the mass-luminosity relation flattens for the
masses just below this threshold. Therefore, 
all stars of mass within a range of several tenths of a solar mass, centered around 
$\rm 2~M_{\odot}$, evolve at similar luminosities during the AGB phase, and, more 
specifically, enter the J region with $\rm M_J \sim -6.4$ mag.

As discussed in the previous section, $\rm \sim 2~M_{\odot}$ stars evolve 
into the J region after experiencing a few TPs following the achievement 
of the C-star stage, and remain within it during a few inter-pulse phases, 
during which the surface carbon and $\rm C/O$ ratio increase from 0.002 to 
0.005 and from $\sim 5$ to $\sim 10$, respectively. At the same time the 
luminosity rises from $\rm 6500~L_{\odot}$ to $\rm 8000~L_{\odot}$. The 
excess of carbon relative to oxygen, which in fact is the key quantity 
driving the formation of dust \citep{fg06}, is of the order of 
$\rm (n_C-n_O)/n_H \sim 3-6 \times 10^{-4}$, which favors the
production of carbonaceous dust (essentially amorphous carbon, given
the low metallicity) with rates spanning the 
$\rm 10^{-10}-10^{-9} ~ M_{\odot}/$yr range. 

We note the steep drop in the JLF at $\rm M_J < -6.4$ mag,
the range of J magnitudes attained by the progeny
of $\rm M > 2~M_{\odot}$ stars. This is partly due to the shorter crossing 
timescale of the J region of these stars than the slightly
lower-mass counterparts, as shown in Fig.~\ref{figtimes}, and also
to the steepening of the mass vs luminosity relationship for the
masses above the threshold for the occurrence of the helium flash.
Indeed the luminosities at which the stars evolve while crossing the J region are $\rm 6-8 \times 10^{3}~L_{\odot}$ for 
$\rm 1.7-2~M_{\odot}$, $\rm 10^4~L_{\odot}$ for $\rm 2.25~M_{\odot}$,
$\rm 1.2 \times 10^4~L_{\odot}$ for $\rm 2.5~M_{\odot}$.

On the faint side of the JLF of Fig.~\ref{fz1m3}
we find the contribution of stars descending from progenitors of
lower mass, the further we move towards the highest $\rm M_J$ values.
The secondary peak in the figure, at $\rm M_J = -5.8$ mag, 
is due to the progeny of $\rm 1~M_{\odot}<M<1.3~M_{\odot}$ stars, 
which evolve within the J region with $\rm -6 < M_J <-5.7$ mag. 
The presence of this secondary peak in the JLF is favored by 
the long duration of the epoch during which these stars formed, 
which extends from $\sim 5$ Gyr to $\sim 2$ Gyr ago (point b above). This
partly counterbalances the shorter staying within the
J region when compared to the higher-mass counterparts
of the same metallicity (point c).

\begin{figure}
\begin{minipage}{0.48\textwidth}
\resizebox{1.\hsize}{!}{\includegraphics{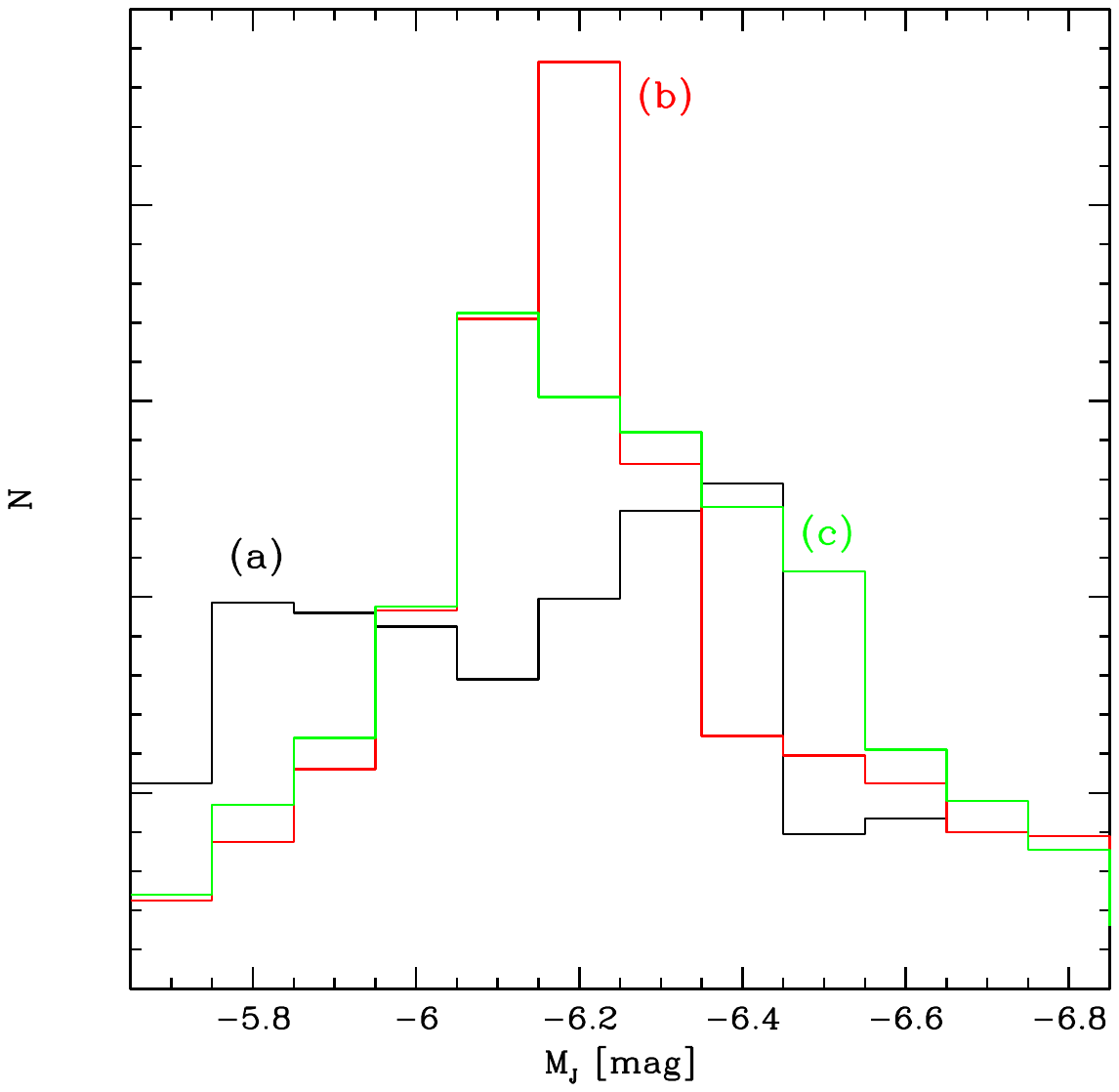}}
\end{minipage}
\vskip-60pt
\caption{The JLF of the stars in the JAGB box
defined by \citet{magnus24}, obtained with three different assumptions
regarding the metal enrichment of the interstellar medium.
(a) - the same case shown in Fig.~\ref{fz1m3}, where no
metal enrichment occurred; (b) - the metal content of the
interstellar medium increases up to $Z=4\times 10^{-3}$, 4 Gyr ago,
and then remained unchanged, (c) - $Z$ increased until $Z=4\times 10^{-3}$, 
5 Gyr ago, then to $Z=8\times 10^{-3}$, 2 Gyr ago.
}
\label{fz8m3}
\end{figure}

Unlike the stars responsible for the primary peak in the JLF,
these sources cross the J region soon after becoming C-stars, with luminosities $\rm \sim 5\times 10^{3}~L_{\odot}$,
and evolve rapidly toward the red side of the CMD. During
this phase their surface carbon is $\sim 2-3 \times 10^{-3}$,
which correspond to $\rm C/O \sim 4-5$ and to a carbon excess
with respect to oxygen $\rm (n_C-n_O)/n_H \sim 2-3\times 10^{-4}$.
The formation rate of carbon dust is of a few $\rm 10^{-10}~M_{\odot}/$yr.

\subsection{The slow metal enrichment}
\label{z4m3}
We now leave the condition of constant metallicity and assume
that the metal content of the galaxy interstellar medium 
increased with different timescales, $\rm \tau_{met}$. 
At this stage we consider only mild metal enrichments,  where the metallicity increases up to $Z=8\times 10^{-3}$.

Fig.~\ref{fz8m3} shows the evolution of the JLF as the timescale of the metal
enrichment gets shorter. For clarity's sake we limit ourselves to showing the 
cases of: a) no metal enrichment, discussed earlier in this section, whose
synthetic distribution on the CMD and the corresponding JLF are shown
in the top, left panel of Fig.~\ref{figmatteo} and in Fig.~\ref{fz1m3}, respectively; 
b) an intermediate case, in which the metallicity gradually increased, until 
reaching $Z=4\times 10^{-3}$, 4 Gyr ago, then remained constant during the 
subsequent epochs; c) the case in which $Z$ increased until $Z=4\times 10^{-3}$, 
5 Gyr ago, then reached $Z=8\times 10^{-3}$, 2 Gyr ago, and has remained 
constant since then. The synthetic CMD obtained in the latter case is shown 
in the bottom, left panel of Fig.~\ref{figmatteo}.

The results reported in Fig.~\ref{fz8m3} indicate that the faster 
the metal enrichment of the interstellar medium, the fainter the peak of the JLF: 
we find $\rm M_J^{peak}=-6.4\ mag, -6.2\ mag, -6.1\ mag$, in the cases (a), (b), (c) listed above, 
respectively. The shape of the JLF is also sensitive
to the timing of the metal enrichment. The two extreme cases are (a), in which most of the JLF is distributed at J fluxes fainter than the peak value, and (c),
in which the majority of the stars populate the $\rm M_J < M_J^{peak}$
tail of the distribution; case (b) is intermediate, as the JLF is 
distributed around the sharp peak, located at $\rm M_J^{peak}=-6.2$ mag.
While $\rm M_J^{peak}$ changes, the mean J magnitude remains
constant at $\rm M_J^{av} \sim -6.2$ mag. The metal enrichment
of the interstellar medium affects the morphology of the JLF, in such a 
way that we gradually move from the $\rm M_J^{av} > M_J^{peak}$ to
the $\rm M_J^{av} < M_J^{peak}$ condition: the difference $\rm (M_J^{peak}-M_J^{av})$ 
is 0.14 mag, 0 mag, -0.14 mag in the cases (a), (b), (c), respectively.

The change in the morphology of the JLF with the timing of the metal
enrichment can be understood by comparing the properties of the stars
of similar mass and different metallicity, in particular in the
age range $1-6$ Gyr. On general grounds, if the metal enrichment occurs earlier than 
$\sim 1$ Gyr ago, the metal-poor stars descending from $\rm \sim 2~M_{\odot}$
progenitors are replaced by their counterparts of similar
mass and higher-metallicity, which, as can be seen in Fig.~\ref{figcmd}, 
are fainter. Under these conditions the peak of the JLF is inevitably 
shifted to lower J fluxes. This is confirmed
by the differences between the JLFs of the cases (a) and (b) shown in 
Fig.~\ref{fz8m3}, which peak at $\rm M_J^{peak}=-6.4$ mag and $\rm M_J^{peak}=-6.2$ mag,
respectively. The reason why the peak of the JLF in the case (b) is sharper than in case (a)
is that both $\rm \sim 2~M_{\odot}$ stars of metallicity $Z \sim 3-4 \times 10^{-3}$
and metal-poor stars of lower mass evolve at $\rm M_J^{peak} \sim -6.2$ mag within 
the J region. No secondary peak is found in
the JLF corresponding to the case (b), because a non negligible fraction of 
the metal-poor, low-mass stars, which we had seen to be responsible for the appearance of the 
secondary peak characterizing the JLF of case (a), are replaced by their higher-metallicity counterparts, which barely enter the J region (see Fig.~\ref{figtimes}).  
In case (c) the peak of the JLF occurs at even fainter magnitudes than in the
cases (a) and (b), because the stars 
spending the longest time within the J region descend from $\rm 1.7-2~M_{\odot}$
progenitors of metallicity $Z = 8 \times 10^{-3}$, which are on average
fainter than their lower-metallicity counterparts. In this case the stars contributing to
the formation of the primary peak at $\rm M_J = -6.1$ mag are a mix of the 
aforementioned $\rm 1.7-2~M_{\odot}$ sources of metallicity $Z = 8 \times 10^{-3}$ and of $\rm \sim 1.5~M_{\odot}$, $Z = 4 \times 10^{-3}$ stars. 

\begin{figure}
\begin{minipage}{0.48\textwidth}
\resizebox{1.\hsize}{!}{\includegraphics{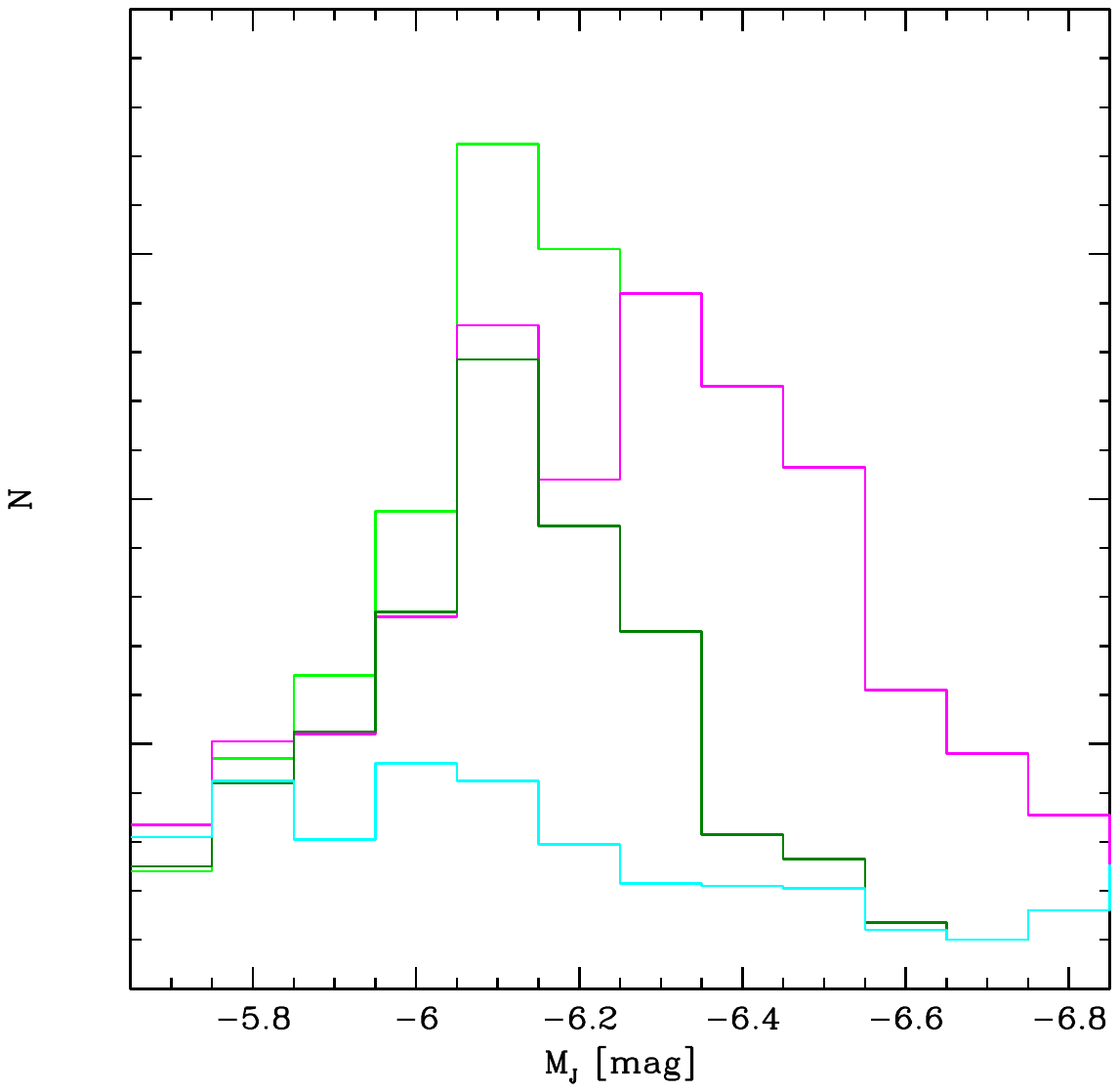}}
\end{minipage}
\vskip-60pt
\caption{JLF of the AGB stars in the
box of the CMD defined by \citet{magnus24}, for different
assumptions regarding the metal enrichment of the interstellar
medium of the galaxy. The color coding is the following:
light green - same as case (c) in Fig.~\ref{fz8m3};
magenta - the metallicity increased up to $Z = 8 \times 10^{-3}$,
4 Gyr ago, then constant after that; dark green - solar metallicity
took over, 1 Gyr ago; cyan - solar metallicity
took over, 2 Gyr ago.
}
\label{fzsun}
\end{figure}

\subsection{Fast metal enrichment, up to solar metallicities}
We now study the shape of the JLF in cases in which the metal 
enrichment proceeded at rates higher than those considered so far, 
up to the most extreme cases, in which solar metallicity has been reached. 
In Fig.~\ref{fzsun} we compare the JLF corresponding to the case (c) in 
Fig.~\ref{fz8m3} with those obtained by assuming that the metallicity 
increased until reaching $Z = 8 \times 10^{-3}$, 4 Gyr ago (magenta track), 
and then with the two most extreme cases, in which the solar metallicity 
was eventually reached, either 1 Gyr ago (dark green line), or 2 Gyr ago 
(cyan line, the corresponding synthetic CMD being shown in the 
bottom, right panel of Fig.~\ref{figmatteo}). 

We found earlier in this section that the faster the metal enrichment
of the interstellar medium, the fainter the $\rm M_J^{\rm peak}$. 
Fig.~\ref{fzsun} shows that this trend is reversed when 
even faster metal enrichments are considered. The comparison between
the case (c) explored previously, when the metallicity $Z = 8 \times 10^{-3}$ was
reached 2 Gyr ago, with the case that the same $Z$ was
reached during earlier epochs, i.e. 4 Gyr ago, shows that in the
latter case the JLF is more distributed towards the bright side, 
with $\rm M_J^{\rm peak}=-6.3$ mag. This difference can be explained by comparing the 
slopes of the $\rm \tau_{JAGB}$ vs mass (or time) trends shown in 
Fig.~\ref{figtimes}, where it is clear that the 
$Z = 8 \times 10^{-3}$ relation is steeper than 
those of the lower metallicities: under these conditions the JAGB 
population is primarily composed of $\rm M \geq 1.5~M_{\odot}$ stars,
the low-mass tail is scarcely present; therefore the JLF peaks at the
magnitudes of the brightest stars. The JLF of the case in question,
shown in magenta in Fig.~\ref{fzsun}, is characterized by the presence of a 
secondary peak, at $\rm M_J=-6.1$ mag, associated with $\rm M < 1.5~M_{\odot}$ 
stars of lower metallicity. While $\rm M_J^{peak}$ changes, 
the mean value is only slightly brighter than in the previous
cases, with $\rm M_J^{av}=-6.24$ mag.

When the solar metallicity takes over, the framework described so far 
changes significantly, owing to the peculiar behavior of the stars of solar
chemistry, which cross the J region much faster than the
sub-solar and metal-poor counterparts (see Fig.~\ref{figtimes}).
If the solar metallicity was reached in epochs more recent
than $\sim 6$ Gyr ago, the population of the
J region is dominated by old stars of sub-solar metallicity,
with a minor, if not negligible, contribution from the younger, solar metallicity stars.
In the case that solar metallicity stars formed only 
during the last 1 Gyr, most of the population of the J
region is composed of $\rm 1.5-2~M_{\odot}$ stars of sub-solar
metallicity: the JLF, shown by the dark green line in
Fig.~\ref{fzsun}, peaks at -6.1 mag. In this case $\rm M_J^{av}=-6.15$ mag.

\begin{figure*}
\vskip-40pt
\begin{minipage}{0.33\textwidth}
\resizebox{1.\hsize}{!}{\includegraphics{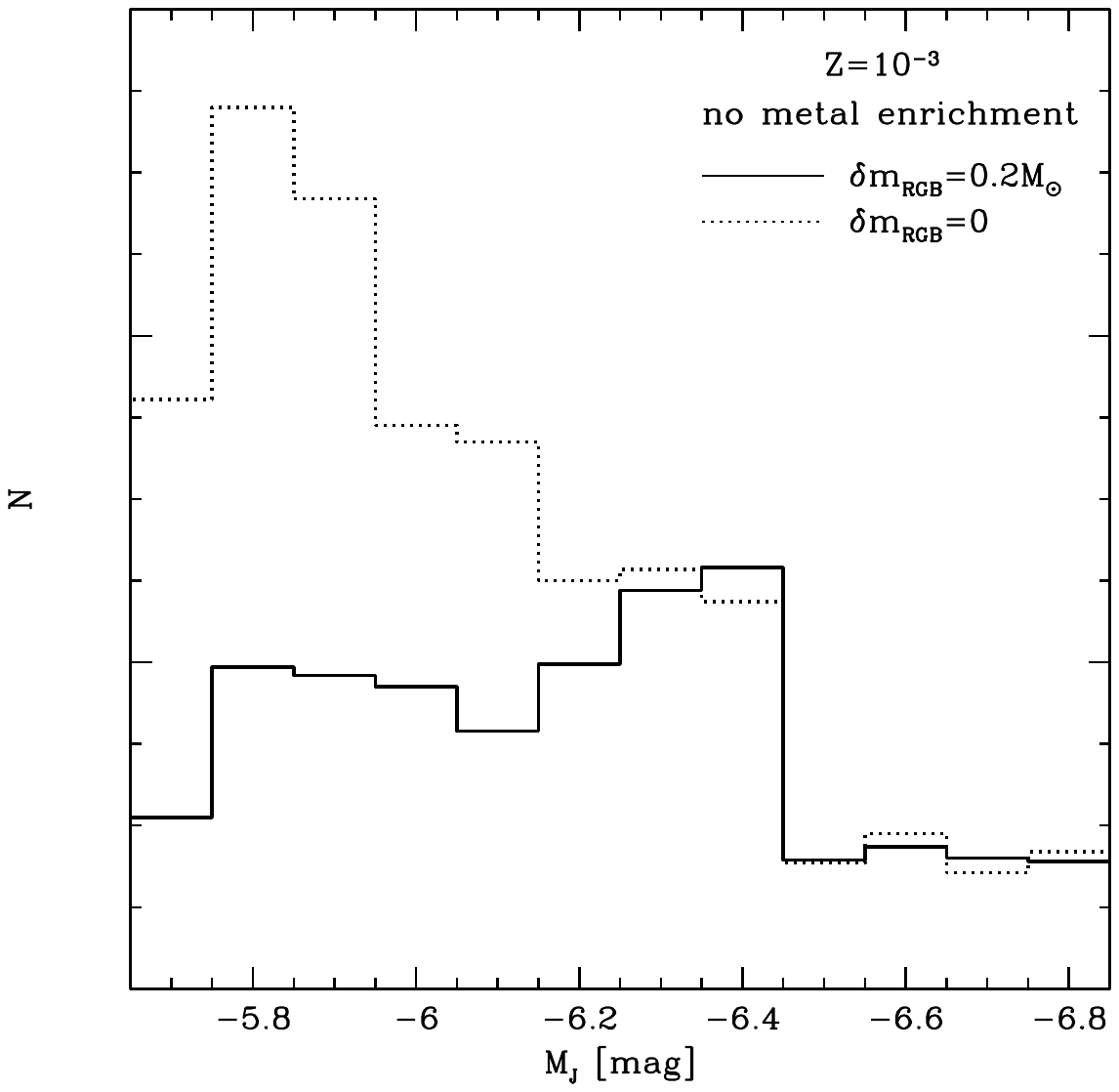}}
\end{minipage}
\begin{minipage}{0.33\textwidth}
\resizebox{1.\hsize}{!}{\includegraphics{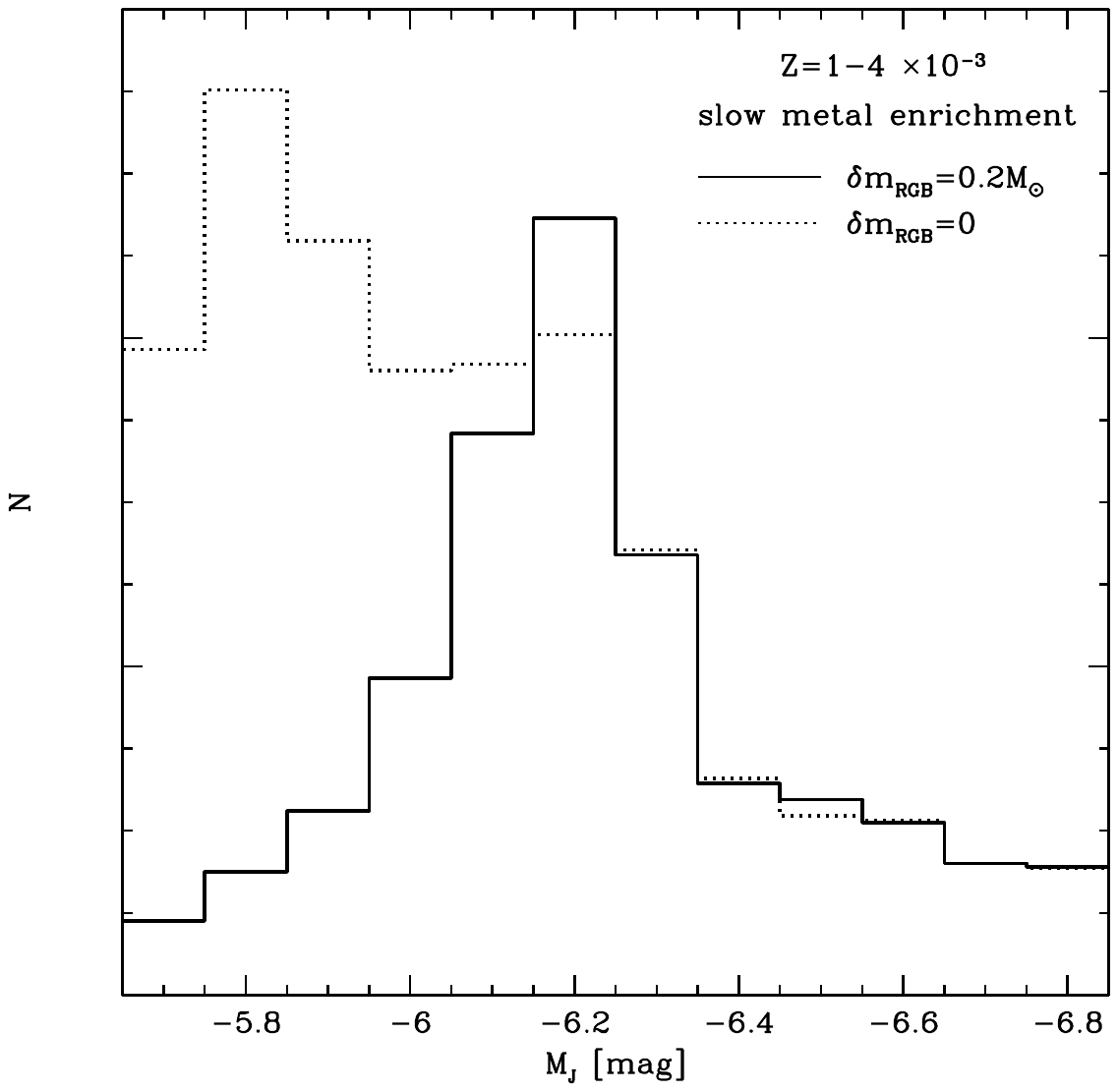}}
\end{minipage}
\begin{minipage}{0.33\textwidth}
\resizebox{1.\hsize}{!}{\includegraphics{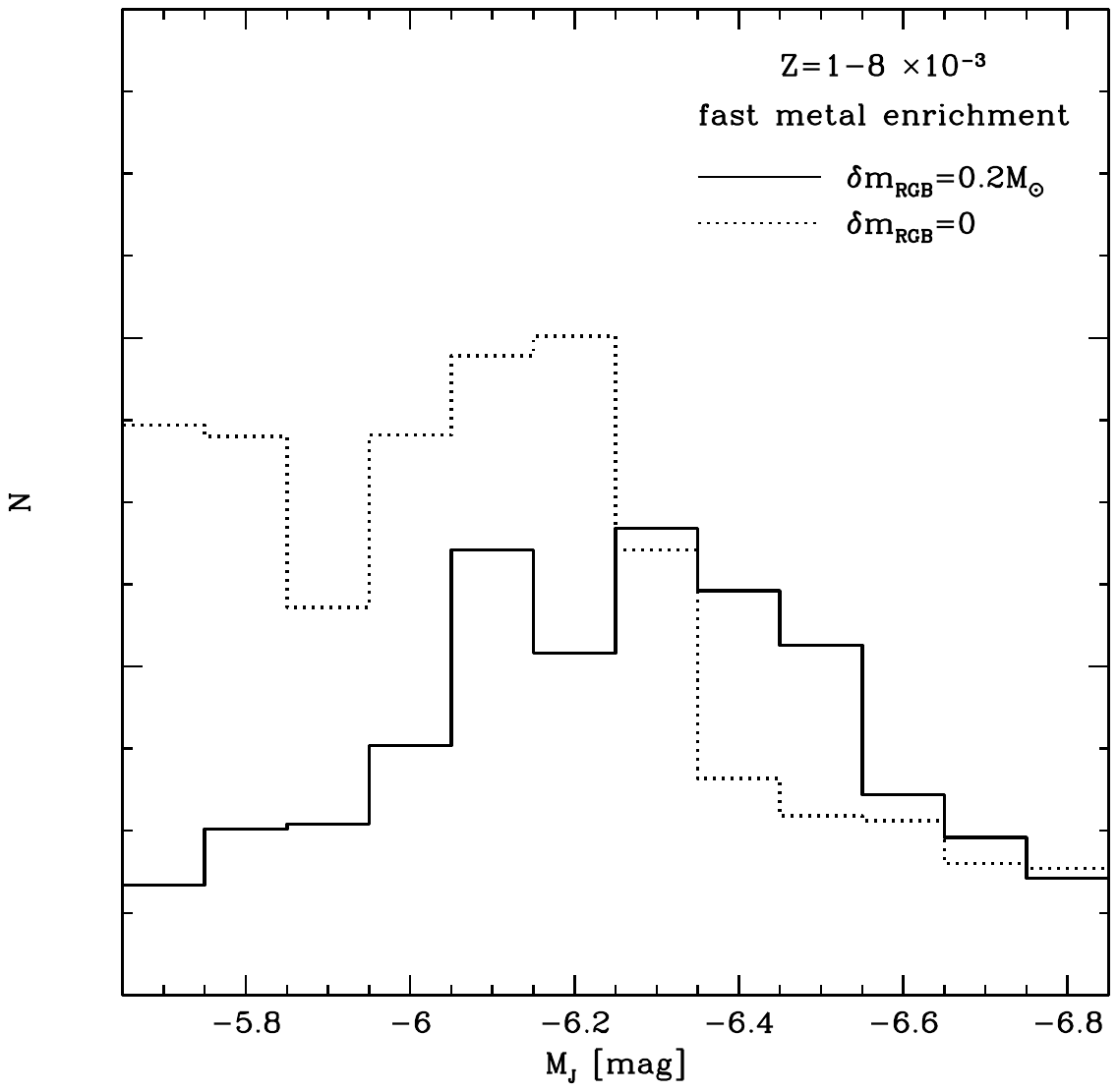}}
\end{minipage}
\vskip-40pt
\caption{Comparison between the JLF obtained by assuming an
average RGB mass loss of $\rm 0.2~M_{\odot}$ (solid track)
with those obtained by assuming that no mass loss occurred
during the RGB phases. The cases presented are the following: 
no metal enrichment and constant metallicity $Z=10^{-3}$
(left panel); metal enrichment up to $Z=4\times 10^{-3}$ (middle panel)
and $Z=8\times 10^{-3}$ (right panel). 
} 
\label{figmloss}
\end{figure*}

\subsection{The role of the mass loss during the red giant branch phase}
\label{mloss}
As discussed in Section \ref{input}, the results presented in this section
are based on the assumption that low-mass stars experienced mass loss
during the RGB evolution, such that on average the mass of the
envelope was reduced by $\rm 0.2~M_{\odot}$ by the time that they reached
the TRGB.
The role of RGB mass loss in the AGB evolution can be understood by noting 
that stars losing mass during the ascent of the RGB start the core helium-burning 
phase and the subsequent AGB evolution with a smaller envelope mass: the AGB 
phase is consequently shorter, since the envelope is lost after the star 
experiences fewer TPs than in the case that no RGB mass loss occurs.
This is mostly relevant for low-mass stars with $\rm M < 1.5~M_{\odot}$,
because these sources evolve longer through the RGB, they suffer
higher mass loss than the more massive counterparts, and even
a few tenths of solar masses lost represent a significant fraction
of the total mass of the envelope.

The RGB mass loss inevitably reflects on the numerical consistency and
the luminosity distribution of the stars populating the J region,
since low-mass stars enter the J region soon after becoming C-stars.
As this condition requires the occurrence of several TDU events, strong 
RGB mass loss prevents low-mass stars from reaching the J region, which 
would then be populated only by stars descending from progenitors of mass 
above solar, so that the JLF would be shifted towards the brighter J fluxes.

To understand the role played by the assumptions regarding the RGB
mass loss on the results obtained by the present analysis, we reconsider here some of the cases discussed in the previous sections
and repeated the simulations based on the population synthesis
approach described in Section \ref{input}, assuming that low-mass 
stars experienced no mass loss during the RGB phase. 
We addressed in particular: the case of no metal enrichment,
presented in Fig.~\ref{fz1m3}; the case with mild metal enrichment,
in which the metallicity of the interstellar medium increases up to
$Z=4\times 10^{-3}$, 4 Gyr ago (the one represented in red in
Fig.~\ref{fz8m3}); the case presented in magenta in Fig.~\ref{fzsun},
where the metallicity increased until reaching $Z=8\times 10^{-3}$,
4 Gyr ago. 

Fig.~\ref{figmloss} presents the comparison between the results
obtained with the standard assumption $\rm \delta m_{RGB}=0.2~M_{\odot}$
with those obtained with no RGB mass loss, for the three
cases considered. In the metal-poor case, the synthetic distribution of
the stars on the CMD obtained in the case that no mass loss 
takes place during the RGB evolution is reported in the top, right
panel of Fig.~\ref{figmatteo}. The corresponding JLF, shown in the left panel
of Fig.~\ref{figmloss}, exhibits a primary peak
at $\rm M_J \sim -5.85$ mag, significantly fainter than the peak 
found in the standard case, located at $\rm M_J \sim -6.4$ mag. This
difference is due to the evolution of the stars descending from
$\rm 0.9-1~M_{\odot}$ progenitors, formed between 9 and 6 Gyr ago:
these stars evolve into the J region if $\rm \delta m_{RGB}=0$,
whereas if their envelope is eroded by the RGB mass 
loss they fail to become C-stars, and thus evolve along the blue
side of the CMD, without entering the J region. These
stars outnumber the higher-mass counterparts, because of the
shape of the IMF, which favors the presence of the stars of lower mass,
and, more important, the duration of the time during which they 
formed, which extends over $\sim 3$ Gyr.

The JLF of the galaxy is heavily affected by the assumptions regarding
the RGB mass loss even in the case that some metal enrichment occurred, such
as the second one considered here. This can be seen in the middle 
panel of Fig.~\ref{figmloss}, where the sharp distribution around the 
peak at $\rm M_J \sim -6.3$ mag, found in the standard case, is replaced
by a double-peaked distribution, in which the primary peak, similarly
to the previous case, is at $\rm M_J \sim -5.85$ mag. This result is not
surprising, considering that the old population of the galaxy is still
made up of low-mass, metal-poor stars, for which the same arguments
exposed previously hold. The primary peak of the JLF would be once more 
due to the presence within the J region of the CMD of the progeny
of $\rm 0.9-1~M_{\odot}$ stars, which have just reached the C-star stage.

We end up with the case of an earlier metal enrichment, which leads
to the formation of $Z=8\times 10^{-3}$ stars in the epochs running from
4 Gyr ago until now. In this case, as shown in the right panel of
Fig.~\ref{figmloss}, the change in the JLF is less dramatic
than in the two previous examples, because the location of the primary peak 
changes by 0.1 mag only if the RGB mass loss is neglected. 
However, there remain notable differences, particularly in the general
shape of the JLF, which is more biased towards the faint side,
when no RGB mass loss is considered. This faint tail of the JLF is due to the
presence in the J region of low-mass, metal-poor stars, formed 
$\sim 8$ Gyr ago.

These findings show that if the mass loss suffered by low-mass stars during the RGB phase
is neglected, the results presented earlier in this section would need
severe revision, particularly in the cases of slow metal enrichment.
Indeed the mean J magnitude of the JAGB population would
be $\rm M_J^{av} \sim -6.05$ mag (to be compared with the $\rm M_J^{av} \sim -6.2$ mag
values found before), which reflects the shift of the
entire JLF to the faint side. When a faster metal enrichment is
considered, only part of the metal-poor, low mass stars enter the J
region, so that $\rm M_J^{av} \sim -6.1$ mag, more similar to the values obtained by assuming $\rm \delta m_{RGB}=0.2~M_{\odot}$. 

While it is important to warn the reader against this source
of uncertainty, we believe that conclusions drawn in the previous
sections are substantially robust, because the mass loss suffered
by stars evolving through the RGB phase is a consolidated result
of the studies of the stellar populations. Besides the aforementioned
investigations on globular clusters, we mentioned e.g. the 
recent study on the evolved stellar population of
Andromeda, by \citet{cla25}, who demonstrated that full consistency
between the results from synthetic modeling and the observed 
distribution of the stars along the CMD built with HST filters
is obtained only if we assume that low-mass stars suffer an average mass
loss of $\rm 0.2~M_{\odot}$ during the RGB evolution. The same
conclusion was reached by Ventura et al. (2026, submitted to A\&A), 
in a study focused on the statistics of RGB and AGB stars in metal-poor 
dwarf and irregular galaxies.

\begin{figure*}
\vskip-40pt
\begin{minipage}{0.48\textwidth}
\resizebox{1.\hsize}{!}{\includegraphics{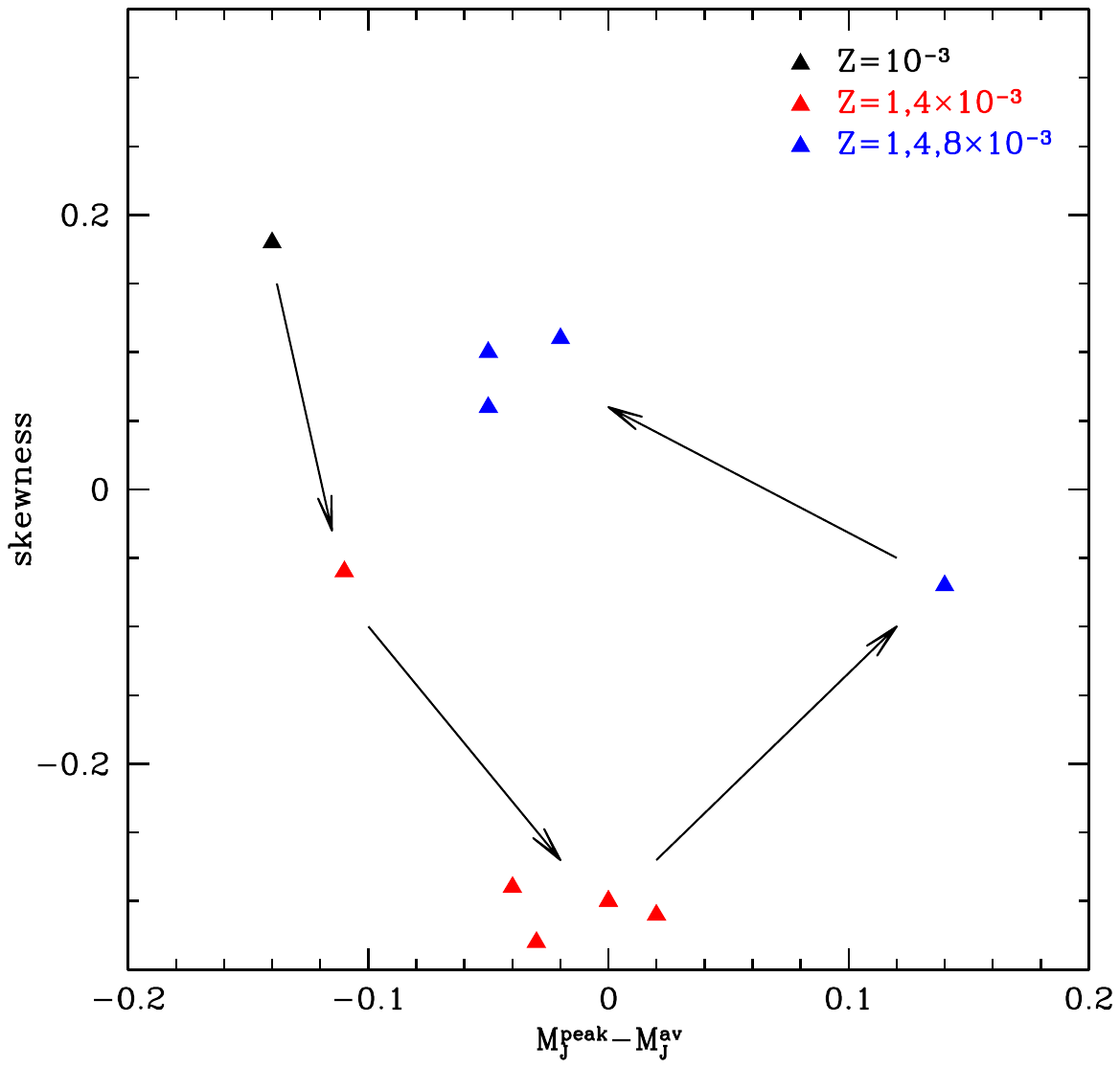}}
\end{minipage}
\begin{minipage}{0.48\textwidth}
\resizebox{1.\hsize}{!}{\includegraphics{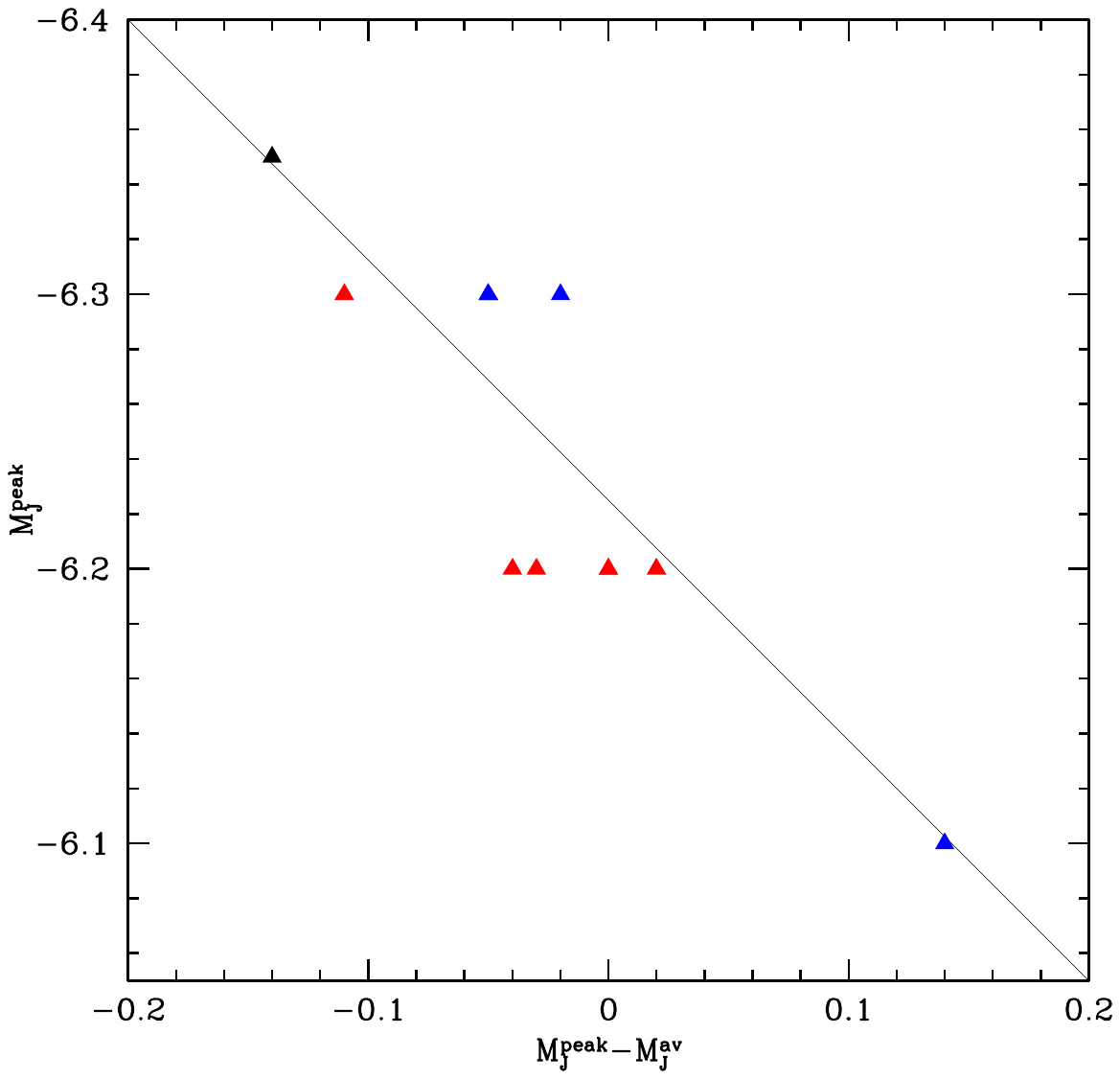}}
\end{minipage}
\vskip-60pt
\caption{Skewness (left panel) and peak J magnitude (right panel) of JLFs of the synthetic stellar populations of the J region
of the $\rm (J-K_S, J)$ diagram, as a function of the difference between the mean 
and the peak J magnitudes. The various points refer to results obtained
with different assumptions regarding the metal enrichment, including
the case where no metal enrichment occurred (black triangle), a few
cases where the final metallicity of the galaxy grew to $Z=4\times 10^{-3}$
(red triangles), and some simulations where the final metallicity of the
interstellar medium is $Z=8\times 10^{-3}$.
} 
\label{figsum}
\end{figure*} 

\section{Discussion}
\label{disc}
The study of the AGB population of the J region of the
CMD is crucial to assess whether these stars can be used as distance 
indicators. This is possible only if we can identify at least some of 
the properties of their LF that are independent of 
the previous history of the galaxies considered. This is a necessary condition for a general application of this method, which
would be otherwise restricted to the galaxies for which the SFH was deduced on the basis of independent investigations. In this
regard, both the mean and the peak J magnitude of JAGB stars
have been investigated as possible candidates for an independent and robust
estimate of galaxy distances
\citep{Lee23,Lee24a,Lee25,Li24,Li25}.

The results presented and discussed in the previous section outlined
the important role played by the rate of metal enrichment
of the interstellar medium on the determination 
of the JLF of the stars populating the J region of a given galaxy. 
The timing of the metal enrichment was shown 
to affect not only the primary parameters of the JLF, such as the peak 
J magnitude $\rm M_J^{peak}$, but more generally the shape of the 
distribution and the details of how the JLF declines 
towards the fainter and the brighter ends.

Before entering the discussion regarding
the sensitivity of the peak and mean J magnitudes of the
stars in the J region to the timing of the metal enrichment of 
the interstellar medium, we deem important to underline that both
quantities are substantially independent of the vertical 
extension (in magnitude) of the JAGB box. This is obvious for $\rm M_J^{peak}$,
as in all the cases investigated in the previous session we found
that it definitively falls within the $(-6.35 - 6.1)$ mag range. 
On the other hand even $\rm M_J^{av}$ would be scarcely affected 
by the size of the JAGB box. Indeed
if the box extends brighter than the upper limit proposed by 
\citet{magnus24}, it is possible that massive AGB stars of mass 
$\rm M \geq 3~M_{\odot}$ enter the J region while experiencing 
HBB. This would contaminate the J region with oxygen-rich objects, 
but their evolutionary times are so short (see e.g. the significant
drop in the $\rm \tau_J$ vs $\rm M$ relation for masses above $\rm 2.5~M_{\odot}$ 
represented by the cyan line in the right panel of Fig.~\ref{figtimes}) 
that only a few of them 
are expected to be counted in, so the estimated mean J magnitude 
would be practically unchanged. This is confirmed by the various cases
examined in the previous session, presented in Figg.~\ref{fz1m3}, 
\ref{fz8m3}, \ref{fzsun}, where it is clear the significant 
drop of the JLF on the bright side of the distribution. 
Even in the case that the lower limit of the JAGB box is fainter than
that suggested by \citet{magnus24} would barely affect $\rm M_J^{av}$:
indeed in this case the stellar population in the JAGB region would
remain substantially unchanged, because the stars of mass below 
$\rm \sim 0.9~M_{\odot}$, which during the AGB phase 
evolve at magnitudes $\rm M_J > -5.7$ mag, do not produce dust, thus
the corresponding evolutionary tracks remain on the blue side of 
the CMD, never entering the JAGB box.

A summary of the results obtained by the various simulations done
in the present work is reported in Fig.~\ref{figsum}, showing how the
skewness of the JLF and $\rm M_J^{peak}$ changes with 
$\rm (M_J^{av}-M_J^{peak})$, the difference between the mean and 
the peak $\rm M_J$. In both planes the curves trace counterclockwise 
trajectories, starting from the case that no metal enrichment 
occurred (Fig.~\ref{fz1m3}), to end up with the case that 
the $Z=8\times 10^{-3}$ metallicity was reached 4 Gyr ago 
(magenta line in Fig.~\ref{fzsun}). 

The initial part of the trends shown in Fig.~\ref{figsum}
extends from the metal-poor galaxy that experienced no
metal enrichment, discussed in Section \ref{z1m3}, to the case 
where the metallicity of the
interstellar medium reached $Z=8\times 10^{-3}$, 2 Gyr ago
(this is shown in light green in Fig.~\ref{fz8m3} and \ref{fzsun}).
In these cases the main factor affecting the change in the 
JLF, discussed in Section \ref{jlf}, is that higher-metallicity stars 
evolve at fainter luminosities than the lower-metallicity 
counterparts of similar mass, so that the peak of the JLF 
occurs at fainter J fluxes as the metal enrichment becomes faster: this is the reason why $\rm (M_J^{peak}-M_J^{av})$ increases from 
-0.15 mag to 0.15 mag, while $\rm M_J^{peak}$ changes from -6.4 mag  to -6.1 mag. 
The shape of the JLF is also sensitive to the metal enrichment, 
as witnessed by the change in the skewness of the distribution.
Initially the JLF (see Fig.~\ref{fz1m3}) is mainly 
distributed on the bright side, thus the skewness is positive
and equal to $\sim 0.2$. The enrichment of metals leads to increasingly asymmetric JLFs towards faint J fluxes (see
Fig.~\ref{fz8m3}), and the value of the skewness eventually 
reaches $\sim -0.3$. The most extreme case, in which the metallicity 
reaches $Z=8\times 10^{-3}$, is the only exception to this trend, 
since the distribution becomes more symmetric, thus the skewness is 
approximately null.

Both the trends reported in Fig.~\ref{figsum} are seen to 
reverse at $\rm (M_J^{peak}-M_J^{av}) \sim 0.15$ mag. From this 
point the most relevant factor for the shape of the JLF is 
that low-mass stars of metallicity $Z \geq 8\times 10^{-3}$ 
barely enter the J region, thus the JLF is determined by the 
progeny of $\rm M \geq 1.5~M_{\odot}$ stars only, which causes 
the LF to move again towards the bright J 
magnitudes: $\rm M_J^{peak}$ decreases 
to -6.3 mag, while $\rm (M_J^{peak}-M_J^{av})$ decreases down to
-0.05 mag, and the skewness of the distribution becomes positive again although small values, of the order of 0.05.

These results indicate that the mean J magnitude of
the JAGB population is  
practically unaffected by the details of the metal enrichment
of the galaxy, and is equal to $\rm M_J^{av} = -6.2$ mag, with 
an uncertainty limited to 0.05 mag. In the majority of cases
the peak J magnitude of the distribution is also 
$\rm M_J^{peak} \sim -6.2$ mag, but a few exceptions,
where it changes by $\sim 0.15$ mag. We note that the data
collected so far, obtained in particular in the massive 
compilations presented by \citet{freedman20} and \citet{freedman25}, 
seem to rule out the occurrence of such extreme cases,
pointing in favour of a universal gaussian distribution,
peaking at $\sim -6.2$ mag. In the evaluation of the quality of
$\rm M_J^{av}$ and $\rm M_J^{peak}$ as distance
indicators of galaxies, it is important to note that while
the mean J magnitude appears more stable and less sensitive
to the metal enrichment of the galaxies, $\rm M_J^{peak}$
has the advantage of being unaffected by possible contamination
by e.g. unresolved, red background galaxies, or even 
faint objects, erroneously scattered into the color range of the 
JAGB stars.

The above arguments indicate that the peak J magnitude
of the JLF cannot be considered as a reliable distance indicator,
because it is sensitive to differences in the metal enrichment histories of galaxies, with variations exceeding 0.3 mag. On the other hand, 
it is remarkable that the changes in $\rm M_J^{peak}$ are
accompanied by variations in the general shape of the JLF, in such
a way that the mean J magnitude of the JAGB population is almost
unaffected by the details of the metal enrichment, and is
equal to $\rm M_J^{av} = -6.2$ mag, with an uncertainty limited
to 0.05 mag.

The use of the J-method to estimate distances in galaxies that experienced very rapid metal enrichment, so that eventually 
solar metallicities are reached, is troublesome. The reason is
that the evolutionary timescales of these objects become
extremely fast after they become C-stars, so the transits
through the J region of the CMD are very fast. If the metal
enrichment was so fast that the solar metallicity was reached 
earlier than $\sim 2$ Gyr ago, the J region would be
populated by low-mass stars of sub-solar metallicity
and the statistics discussed so far would no longer hold.

\section{Conclusions}
\label{concl}
We used population synthesis results to investigate the
applicability of the J method to derive the distances of 
galaxies. We assumed a constant SFR form galaxy formation to present and we focused on the effects of the timing of  interstellar medium enrichment on the shape of the JLF for the stars in the J region of the CMD. 

The stars currently populating the J region descend from
progenitors of mass in the $\rm 1-3~M_{\odot}$ range, formed
during the epochs ranging from 6 Gyr to around 1 Gyr ago.
Indeed older stars barely enter the J region, whereas
the stars younger than $\sim 1$ Gyr evolve brighter than the
upper limit of the box selected to define the J region,
and in any case their crossing time would be too fast to 
meaningfully affect the statistics. The progeny of
$\rm 2-2.5~M_{\odot}$ stars provides the most relevant 
contribution to the JAGB population, 
because they evolve within the J region for a longer
period than the lower-mass counterparts.

The shape of the JLF turns out to be extremely sensitive
to the timing of the metal enrichment of the galaxy, owing to the 
differences between the evolutionary properties of stars of 
similar mass and different metallicity. This concerns in particular
the time evolution of the luminosity and the time when
the C-star stage is reached, both of which are relevant for the shape of the JLF
and the peak of the distribution of the J magnitudes
of the stars in the J region.

The peak J magnitude of the JLF cannot be used to infer
the distance of galaxies, because it is extremely sensitive to the
metal enrichment process, and changes by 0.3 mag, spanning the range 
from $\rm M_J^{peak} = -6.4$ mag to $\rm M_J^{peak} = -6.1$ mag. 
On the other hand, the mean 
J magnitude turns out to be a much more robust distance indicator, 
as it is found to be $\rm M_J^{av} = -6.2 \pm 0.05$ mag, 
independently of the 
timing of the metal enrichment process.

The applicability of the method to the galaxies where the metal 
enrichment was so fast that solar metallicity stars older than 
$\sim 2$ Gyr formed is questionable, because these stars evolve very fast 
during the evolutionary phases following the achievement 
of the C-star stage, so that their transit across the J
region of the CMD is extremely rapid.

The applicability of the J method to measure distances
must be further tested against variations in the star
formation rate, which alter the relative fractions of
the stars of different age and mass, which might in turn lead to changes in the mass and luminosity distribution of the stars populating
the J region.

%

\begin{thebibliography}{}
\bibitem[Bl\"ocker \& Sch\"onberner(1991)]{blo91} Bl\"ocker, T. \& Sch\"onberner, D.\ 1991, \aap, 244, L43

\bibitem[\protect\citeauthoryear{Bortolini et al.}{2024}]{bortolini24} Bortolini G., {\"O}stlin G., Habel N., Hirschauer A.~S., Jones O.~C., Justtanont K., Meixner M., et al., 2024, A\&A, 689, A146. 

\bibitem[Bortolini et al.(2025)]{bortolini25} Bortolini, G., Correnti, M., Adamo, A., et al.\ 2025, \apj, 991, 2, 212

\bibitem[Correnti et al.(2025)]{correnti25} Correnti, M., Bortolini, G., Dell'Agli, F., et al.\ 2025, \apj, 990, 1, 72

\bibitem[Dell'Agli et al.(2014a)]{flavia14a} Dell'Agli, F., Garc{\'\i}a-Hern{\'a}ndez, D.~A., Rossi, C., et al.\ 2014a, \mnras, 441, 1115.

\bibitem[Dell'Agli et al.(2014)]{flavia14b} Dell'Agli, F., Ventura, P., Garcia Hernandez, D.~A., et al.\ 2014, \mnras, 442, L38

\bibitem[Dell'Agli et al.(2015a)]{flavia15a} Dell'Agli, F., Ventura, P., Schneider, R., et al.\ 2015a, \mnras, 447, 2992

\bibitem[Dell'Agli et al.(2015b)]{flavia15b} Dell'Agli, F., Garc{\'\i}a-Hern{\'a}ndez, D.~A., Ventura, P., et al.\ 2015b, \mnras, 454, 4, 4235

\bibitem[Dell'Agli et al.(2016)]{flavia16} Dell'Agli, F., Di Criscienzo, M., Boyer, M.~L., et al.\ 2016, \mnras, 460, 4, 4230

\bibitem[Dell'Agli et al.(2018)]{flavia18} Dell'Agli, F., Di Criscienzo, M., Ventura, P., et al.\ 2018, \mnras, 479, 5035

\bibitem[Dell'Agli et al.(2019)]{flavia19} Dell'Agli, F., Di Criscienzo, M., Garc{\'\i}a-Hern{\'a}ndez, D.~A., et al.\ 2019, \mnras, 482, 4, 4733.

\bibitem[Ferrarotti \& Gail(2002)]{fg02} Ferrarotti, A.~S. \& Gail, H.-P.\ 2002, \aap, 382, 256

\bibitem[Ferrarotti \& Gail(2006)]{fg06} Ferrarotti, A.~S. \& Gail, H.-P.\ 2006, \aap, 447, 553

\bibitem[Freedman \& Madore(2020)]{freedman20} Freedman, W.~L. \& Madore, B.~F.\ 2020, \apj, 899, 1, 67

\bibitem[\protect\citeauthoryear{Freedman et al.}{2025}]{freedman25} Freedman W.~L., Madore B.~F., Hoyt T.~J., Jang I.~S., Lee A.~J., Owens K.~A., 2025, ApJ, 985, 203. 

\bibitem[Gavetti et al.(2025)]{cla25} Gavetti, C., Ventura, P., Dell'Agli, F., et al.\ 2025, \aap, 699, A23

\bibitem[Gavetti et al.(2026)]{cla26} Gavetti, C., Ventura, P., Dell'Agli, F., et al.\ 2026, arXiv:2601.19594

\bibitem[Harmsen et al.(2023)]{harmsen23} Harmsen, B., Bell, E.~F., D'Souza, R., et al.\ 2023, \mnras, 525, 449 

\bibitem[Kamath et al.(2023)]{devika23} Kamath, D., Dell'Agli, F., Ventura, P., et al.\ 2023, \mnras, 519, 2169

\bibitem[Kobayashi et al.(2020)]{ciaki20} Kobayashi, C., Karakas, A.~I., \& Lugaro, M.\ 2020, \apj, 900, 2, 179
 
\bibitem[\protect\citeauthoryear{Kroupa}{2001}]{kroupa01} Kroupa P., 2001, MNRAS, 322, 231. 

\bibitem[\protect\citeauthoryear{Lee}{2023}]{Lee23} Lee A.~J., 2023, ApJ, 956, 15. 

\bibitem[\protect\citeauthoryear{Lee et al.}{2024a}]{Lee24a} Lee A.~J., Freedman W.~L., Jang I.~S., Madore B.~F., Owens K.~A., 2024, ApJ, 961, 132.

\bibitem[\protect\citeauthoryear{Lee et al.}{2024b}]{Lee24b} Lee, A.~J., Weisz, D.~R., Ren, Y., et al.\ 2024, , arXiv:2410.09256. 

\bibitem[\protect\citeauthoryear{Lee et al.}{2025}]{Lee25} Lee A.~J., Freedman W.~L., Madore B.~F., Jang I.~S., Owens K.~A., Hoyt T.~J., 2025, ApJ, 985, 182. 

\bibitem[\protect\citeauthoryear{Li et al.}{2024}]{Li24} Li S., Riess A.~G., Casertano S., Anand G.~S., Scolnic D.~M., Yuan W., Breuval L., et al., 2024, ApJ, 966, 20

\bibitem[\protect\citeauthoryear{Li et al.}{2025}]{Li25} Li S., Riess A.~G., Scolnic D., Casertano S., Anand G.~S., 2025, ApJ, 988, 97. 

\bibitem[Madore \& Freedman(2020)]{madore20} Madore, B.~F. \& Freedman, W.~L.\ 2020, \apj, 899, 1, 66

\bibitem[Magnus et al.(2024)]{magnus24} Magnus, E., Groenewegen, M.~A.~T., Girardi, L., et al.\ 2024, \aap, 691, A350

\bibitem[Marigo(2002)]{marigo02} Marigo, P.\ 2002, \aap, 387, 507

\bibitem[Marini et al.(2021)]{marini21} Marini, E., Dell'Agli, F., Groenewegen, M.~A.~T., et al.\ 2021, \aap, 647, A69

\bibitem[Mazzitelli(1979)]{italo} Mazzitelli, I.\ 1979, \aap, 79, 1-2, 251

\bibitem[Nanni et al.(2013)]{nanni13} Nanni, A., Bressan, A., Marigo, P., et al.\ 2013, \mnras, 434, 3, 2390

\bibitem[Nanni et al.(2014)]{nanni14} Nanni, A., Bressan, A., Marigo, P., et al.\ 2014, \mnras, 438, 3, 2328

\bibitem[Nanni et al.(2016)]{nanni16} Nanni, A., Marigo, P., Groenewegen, M.~A.~T., et al.\ 2016, \mnras, 462, 2, 1215

\bibitem[Nanni et al.(2018)]{nanni18} Nanni, A., Marigo, P., Girardi, L., et al.\ 2018, \mnras, 473, 4, 5492

\bibitem[Nanni et al.(2019)]{nanni19} Nanni, A., Groenewegen, M.~A.~T., Aringer, B., et al.\ 2019, \mnras, 487, 1, 502

\bibitem[Nenkova et al.(1999)]{nenkova99} Nenkova, M., Ivezic, Z., \& Elitzur, M.\ 1999, 
Thermal Emission Spectroscopy and Analysis of Dust, Disks, and Regoliths, 20

\bibitem[Nikolaev \& Weinberg(2000)]{nikolaev00} Nikolaev, S. \& Weinberg, M.~D.\ 2000, \apj, 542, 2, 804

\bibitem[Ripoche et al.(2020)]{ripoche20} Ripoche, P., Heyl, J., Parada, J., et al.\ 2020, \mnras, 495, 3, 2858

\bibitem[Romano(2022)]{romano22} Romano, D.\ 2022, \aapr, 30, 1, 7.

\bibitem[Sackmann \& Boothroyd(1992)]{sack} Sackmann, I.-J. \& Boothroyd, A.~I.\ 1992, \apjl, 392, L71. doi:10.1086/186428

\bibitem[Schneider \& Maiolino(2024)]{raffa24} Schneider, R. \& Maiolino, R.\ 2024, \aapr, 32, 1, 2

\bibitem[Tailo et al.(2021)]{tailo21} Tailo, M., Milone, A.~P., Lagioia, E.~P., et al.\ 2021, \mnras, 503, 694

\bibitem[Valiante et al.(2009)]{valiante09} Valiante, R., Schneider, R., Bianchi, S., et al.\ 2009, \mnras, 397, 3, 1661

\bibitem[Valiante et al.(2011)]{valiante11} Valiante, R., Schneider, R., Salvadori, S., et al.\ 2011, \mnras, 416, 3, 1916

\bibitem[Ventura et al.(1998)]{ventura98} Ventura, P., Zeppieri, A., Mazzitelli, I.,   D'Antona, F., 1998, A\&A, 334, 953

\bibitem[Ventura et al.(2001)]{ventura01} Ventura, P., D'Antona, F., Mazzitelli, I., et al.\ 2001, \apjl, 550, L65.

\bibitem[Ventura et al.(2012)]{ventura12} Ventura P., Di Criscienzo M., Schneider R., 
et al.\ 2012, \mnras, 420, 1442

\bibitem[Ventura et al.(2014)]{ventura14} Ventura P., Dell'Agli F., Schneider R., et al.\ 2014, \mnras, 439, 977

\bibitem[Ventura et al.(2018)]{ventura18} Ventura, P., Karakas, A., Dell'Agli, F., et  al.\ 2018, \mnras, 475, 2282. 

\bibitem[Ventura et al.(2022)]{ventura22} Ventura, P., Dell'Agli, F., Tailo, M., et al.\ 2022, Universe, 8, 45.

\bibitem[Ventura et al.(2026)]{ventura26} Ventura, P., D'Souza, R., Dell'Agli, F., et al.\ 2026, arXiv:2603.09879.

\bibitem[Mancini et al.(2015)]{mancini15} Mancini, M., Schneider, R., Graziani, L., Valiante, R., Dayal, P., Maio, U., Ciardi, B. and Hunt, L.~K.\ 2015, MNRASL, 451, 70.

\bibitem[Graziani et al.(2020)]{graziani20} Graziani, L., Schneider, R., Ginolfi, M., Hunt, L.~K., Maio, U., Glatzle, M. and Ciardi, B.\ 2015, MNRAS, 494, 1071.

\bibitem[Aoyama et al.(2018)]{aoyama18} Aoyama, S., Hou, Kuan-Chou, Hirashita, H., Nagamine, K., Shimizu, I.\ 2018, MNRAS, 478, 4905.

\bibitem[Ginolfi et al.(2018)]{ginolfi18} Ginolfi, M., Graziani, L., Schneider, R., Marassi, S., Valiante, R., Dell'Agli, F., Ventura, P., Hunt, L.~K.\ 2018, MNRAS, 478, 4905.

\bibitem[Marconi et al.(1995)]{marconi95} Marconi, G., Tosi, M., Greggio, L., Focardi, P.\ 1995, AJ, 109, 173.

\bibitem[Tolstoy et al.(2009)]{TolstoyReview09} Tolstoy, E., Hill, V., Tosi, M.\ 2009, ARAA, 47, 371.

\bibitem[Conroy (2013)]{ConroyReview13} Conroy, C.\ 2013, ARAA, 51, 393.

\bibitem[Madau \& Dickinson (2014)]{MadauDickinsonReview14} Madau, P. and Dickinson, M.\ 2014, ARAA, 52, 415.

\bibitem[Pacifici et al. (2016)]{Pacifici16} Pacifici, C., Kassin, S., Weiner, B. et al.\ 2016, ApJ, 832, 79.

\bibitem[Diemer et al. (2017)]{Diemer17} Diemer, B., Sparre, M., Abramson, L., Torrey, P. \ 2017, ApJ, 839, 26.

\bibitem[Grebel, E.~K. (2000)]{Grebel20} Grebel, E.~K. \ 2020, ESA Special Publication, 445, 87.

\bibitem[Rogers, B. et al. (2010)]{Rogers10} Rogers, B., Ferreras, I.,et al. \ 2010, MNRAS, 402, 447.

\bibitem[{Jegatheesan}, K, et al. (2025)]{Jegatheesan25} Jegatheesan, K.,et al. \ 2025, A\&A, 694, 224.


\bibitem[{Calura}, F, et al. (2004)]{Calura04} Calura, F., Matteucci, F., Menci, N. \ 2004, MNRAS, 353, 500.

\end{thebibliography}
%

\end{document}